\documentclass[twocolumn,tighten,trackchanges]{aastex631}
\PassOptionsToPackage{dvipsnames,svgnames,table}{xcolor}

\usepackage{amsmath}
\usepackage{graphicx}
\usepackage[dvipsnames]{xcolor}
\usepackage{apjfonts}
\usepackage{booktabs}
\usepackage{tablefootnote}
\usepackage{lipsum}
\usepackage{xspace}

\newcommand{\kms}{\,km\,s$^{-1}$}

\def\lsim{\hbox{\rlap{\raise 0.425ex\hbox{$<$}}\lower 0.65ex\hbox{$\sim$}}}
\def\gsim{\hbox{\rlap{\raise 0.425ex\hbox{$>$}}\lower 0.65ex\hbox{$\sim$}}}

\def\arcsec{\hbox{$^{\prime\prime}$}}
\def\arcdeg{\mbox{$^\circ$}}

\newcommand{\CosmosSNIa}{SN~2025ogs\xspace}
\newcommand{\snz}{2.05}

\definecolor{maroon}{rgb}{0.760,0.118,0.337}

\newcommand{\STScI}{Space Telescope Science Institute, Baltimore, MD 21218, USA}
\newcommand{\JHU}{William H. Miller III Department of Physics \& Astronomy, Johns Hopkins University, 3400 N Charles St, Baltimore, MD 21218, USA}

\newcommand{\NEF}{NASA Einstein Fellow}

\usepackage[mathlines]{lineno}

\shorttitle{A Benchmark SN~Ia at $z=2$}
\shortauthors{Siebert et al.}

\graphicspath{{./}{figures/}}

\begin{document}

\title{\CosmosSNIa: A Spectroscopically-Normal Type Ia Supernova at z = 2 as a Benchmark for Redshift Evolution}

\correspondingauthor{Matthew~R.~Siebert}
\email{msiebert@stsci.edu}

\author[0000-0003-2445-3891]{M.~R.~Siebert}
\affiliation{\STScI}

\author[0000-0002-2361-7201]{J.~D.\ R.\ Pierel}
\altaffiliation{\NEF}
\affiliation{\STScI}

\author[0000-0003-0209-674X]{M.~Engesser}
\affiliation{\STScI}

\author[0000-0003-4263-2228]{D.~A.~Coulter} 
\affiliation{\STScI}
\affiliation{William H. Miller III Department of Physics \& Astronomy, Johns Hopkins University, 3400 N Charles St, Baltimore, MD 21218, USA}

\author[0000-0002-4781-9078]{C.~DeCoursey}
\affiliation{Steward Observatory, University of Arizona, 933 N. Cherry Avenue, Tucson, AZ 85721 USA}

\author[0000-0003-2238-1572]{O.~D.~Fox} 
\affiliation{\STScI}

\author[0000-0002-4410-5387]{A.~Rest} 
\affiliation{\STScI}
\affiliation{William H. Miller III Department of Physics \& Astronomy, Johns Hopkins University, 3400 N Charles St, Baltimore, MD 21218, USA}

\author[0000-0003-1060-0723]{W.~Chen} 
\affiliation{Department of Physics,
Oklahoma State University, 145 Physical Sciences Bldg, Stillwater, OK
74078, USA}

\author[0000-0002-7566-6080]{J.~M.~DerKacy} 
\affiliation{\STScI}

\author[0000-0003-1344-9475]{E.~Egami}
\affiliation{Steward Observatory, University of Arizona, 933 N. Cherry Avenue, Tucson, AZ 85721 USA}

\author[0000-0002-2445-5275] {R.~J.~Foley}
\affiliation{Department of Astronomy and Astrophysics, University of California, Santa Cruz, 1156 High Street, Santa Cruz CA 96054, USA}

\author[0000-0002-6230-0151]{D.~O.~Jones} 
\affiliation{Institute for Astronomy, University of Hawai'i, 640 N.~A'ohoku Pl., Hilo, HI 96720, USA}

\author[0000-0002-6610-2048]{K.~Kakiichi} 
\affiliation{Cosmic Dawn Center (DAWN), Denmark}
\affiliation{Niels Bohr Institute, University of Copenhagen, Jagtvej 128, DK-2200 Copenhagen N, Denmark}

\author[0000-0002-6610-2048]{A.~M.~Koekemoer} 
\affiliation{\STScI}

\author[0009-0003-8380-4003]{Z.~G.~Lane} 
\affiliation{School of Physical and Chemical Sciences—Te Kura Matū, University of Canterbury, Private Bag 4800, Christchurch 8140, New Zealand}

\author[0000-0003-2037-4619]{C.~Larison} 
\affiliation{\STScI}

\author[0000-0001-7839-1986]{D.~C.~Leonard} 
\affiliation{Department of Astronomy, San Diego State University, San Diego, CA 92182-1221, USA}

\author[0000-0003-1169-1954]{T.~J.~Moriya}
\affiliation{National Astronomical Observatory of Japan, National Institutes of Natural Sciences, 2-21-1 Osawa, Mitaka, Tokyo 181-8588, Japan}
\affiliation{Graduate Institute for Advanced Studies, SOKENDAI, 2-21-1 Osawa, Mitaka, Tokyo 181-8588, Japan}
\affiliation{School of Physics and Astronomy, Monash University, Clayton, Victoria 3800, Australia}

\author[0000-0003-0209-9246]{E.~Padilla~Gonzalez} 
\affiliation{\JHU}

\author[0000-0001-9171-5236]{R.~M.~Quimby}
\affiliation{Department of Astronomy/Mount Laguna Observatory, San Diego State University, 5500 Campanile Drive, San Diego, CA 92812-1221, USA}
\affiliation{Kavli Institute for the Physics and Mathematics of the Universe (WPI), The University of Tokyo Institutes for Advanced Study,
The University of Tokyo, Kashiwa, Chiba 277-8583, Japan
}

\author[0000-0002-2798-2943]{K.~Shukawa} 
\affiliation{William H. Miller III Department of Physics \& Astronomy, Johns Hopkins University, 3400 N Charles St, Baltimore, MD 21218, USA}

\author[0000-0002-7756-4440]{L.~G.~Strolger} 
\affiliation{\STScI}

\author[0000-0002-0632-8897]{Y. Zenati}
\affiliation{William H. Miller III Department of Physics \& Astronomy, Johns Hopkins University, 3400 N Charles St, Baltimore, MD 21218, USA}
\affiliation{Astrophysics Research Center of the Open University (ARCO), Ra’anana 4353701, Israel}
\affiliation{Department of Natural Sciences, The Open University of Israel, Ra’anana 4353701, Israel}

\begin{abstract}
The \textit{Nancy Grace Roman Space Telescope} will provide a revolutionary measurement of the Universe's expansion kinematics, driven by dark matter and dark energy, out to $z \approx 3$. The accuracy of this measurement is predicated on the assumption that standardized Type Ia supernova (SN~Ia) luminosities do not evolve with redshift. If present, SN~Ia luminosity evolution is expected to be most detectable in the dark matter-dominated era of the Universe ($z \gtrsim 1.5$), with its effects becoming more easily distinguishable from dark energy variation at increasing redshift. We present \textit{JWST} NIRCam and NIRSpec observations of SN~2025ogs, a normal SN~Ia at $z = 2.05 \pm 0.01$. This SN offers a key point of comparison for interpreting future high-redshift SN~Ia samples. The NIRCam light curve indicates a blue color ($B - V = -0.27 \pm 0.06$ mag) and a moderately fast decline ($\Delta m_{15}(B) = 1.55 \pm 0.15$ mag), both within standard criteria for inclusion in cosmological analyses. Its luminosity distance is in $1.0\sigma$ agreement with a standard flat $\Lambda$CDM model, as well as with current cosmological constraints from the Dark Energy Survey (DES 5yr) and Pantheon+. The NIRSpec spectrum displays all of the hallmark absorption features of a normal SN~Ia observed at peak brightness. We find that the rest-frame optical color, rest-frame near-ultraviolet properties, and \ion{Si}{2} line strengths are all consistent with the moderately fast decline inferred from the light curve. Multiple absorption features (\ion{Ca}{2} H\&K, \ion{O}{1} $\lambda7774$, and the \ion{Ca}{2} NIR triplet) all appear at a lower blueshift relative to a sample of low-$z$ SNe~Ia. Together, these results suggest that SN~Ia standardization remains robust at $z \approx 2$, and also highlight the importance of \textit{JWST} spectroscopy for uncovering evolutionary effects that could impact \textit{Roman}'s high-precision cosmology.
\end{abstract}


\section{Introduction}
Type Ia supernovae (SNe~Ia) are luminous enough to be observed over cosmological distances. Well-characterized relationships between their luminosities and their light curve properties \citep{Phillips93,Tripp98} have enabled their use as highly-precise distance indicators, which led to the discovery of the accelerating rate of expansion of the Universe \citep{Riess98:lambda, Perlmutter99}. SNe~Ia are still at the forefront of modern cosmology and understanding their remaining systematic uncertainties is of critical importance to the success missions like the Vera C. Rubin Observatory Legacy Survey of Space and Time~\citep[LSST;][]{LSST_Sci_Col09,LSST_Sci_Col17} and the Roman Space Telescope High Latitude Time Domain Survey \citep[HLTDS;][]{jha25}. Recent results using observations from the Dark Energy Spectroscopic Instrument (DESI; \citealt{Adame25}) show a preference for a cosmological model with an evolving dark energy equation of state ($w_{0}w_{a}$CDM) over a cosmological constant ($\Lambda$CDM). The Rubin LSST and \textit{Roman} HLTDS will measure SN~Ia properties out to $z\sim3$ and will be instrumental for corroborating (or refuting) the existing evidence for evolving dark energy which is currently constrained at the $2.5-4\sigma$ level \citep{Scolnic22, Adame25}. 

Using photometric measurements alone, the scatter in SN~Ia distance measurements can be reduced to $\sim8\%$ \citep{Jones18:ps1, Scolnic22} and further improvements to precision require additional parameters. It is likely that the remaining intrinsic scatter is related to environmental properties or explosions physics \citep{Conley11, Scolnic18:ps1, Siebert20a}. For example, there is a well-studied relationship between SN~Ia Hubble residuals (the difference between measured distance moduli and those inferred from a cosmological model) and host galaxy mass \citep{Kelly10, Lampeitl10:host, Sullivan10}, referred to as the host galaxy ``mass-step''. This relationship is likely a proxy for some physical origin, possibly a property of the progenitor system like metallicity \citep{Moreno-Raya16} or a global property like host-galaxy dust \citep{brout_its_2021} or star formation rate \citep{Martin24,Dixon25}. Regardless of the physical origin, these properties, and SN\,Ia progenitor channels \citep{Maoz14,Childress14}, are expected to evolve with redshift, plausibly causing redshift evolution in SN~Ia properties. 

The potential for SN~Ia luminosity evolution is a concern for next-generation cosmological experiments like the Rubin LSST and Roman HLTDS. If unaccounted SN luminosity evolution is significant, this will introduce a redshift-dependent bias in our SN~Ia distance measurements \citep{Riess06}, which would propagate to inaccurate constraints on cosmological parameters and the potential for dark energy evolution. Therefore, characterizing SNe~Ia at the highest redshifts possible (where evolution effects are strongest and distinct from dark energy evolution; \citealt{Riess06, Pierel24,pierel_testing_2025}) 
will inform any necessary corrections and additional data/analysis needed to improve the robustness and precision of dark energy constraints from next-generation cosmological experiments.


At low- and intermediate-redshifts, the optical spectroscopic diversity of SNe~Ia is incredibly well characterized. There exist a wide variety of well-documented peculiar subclasses of SNe~Ia (e.g., 91T-, 91bg-, 02es-, 03fg-like, Ia-CSM, and Iax; \citealt{Taubenberger17}, references therein) whose relative contribution at high-redshift is not known. Normal-SNe~Ia are more common at low-redshift and have more uniform observational properties, thus making them useful for cosmology. Among normal-SNe~Ia, additional subclasses have been identified that are broadly correlated with SN~Ia luminosity and ejecta velocity \citep{Branch06}. Ejecta velocity is correlated with intrinsic SN~Ia color \citep{Foley11:vel}, and potentially impacts cosmological distance measurements \citep{Siebert20a, Dettman21, Pan24, Burgaz25}. There is now compelling evidence that variations in dust properties (affecting SN~Ia color) add scatter and potential bias to the Hubble residuals \citep{Brout21,Gall24}. Furthermore, SNe~Ia that exhibit unusually large blueshifts of their \ion{Si}{2} 6355 \AA ~feature at early times,  traditionally referred to as high-velocity (HV; $v_{Si}$ $ <-12,000$ \kms), are correlated with their host environments \citep{Foley12:vel, Pan20:vel,Nugent23} and statistically have more circumstellar material (CSM) \citep{Sternberg11}. These are properties that are affected by progenitor age and expected to evolve with redshift. There is some evidence that the fraction of HV SNe~Ia decreases in the redshift range of $z\sim 0.1 - 0.5$ \citep{Pan24}. This effect could be related to metallicity \citep{Pan20:vel}, and thus, we might expect this trend to continue at even higher-redshifts. Cosmologically-useful SNe span a large range of this parameter space and each subclass (aside from the most extreme SNe) is represented in our distance measurements. Understanding the relative contributions of these subclasses in high-$z$ samples is an important first step toward constraining how SN~Ia progenitors might be changing with redshift. If significant luminosity evolution is confirmed, this data will help to understand its physical origin. 

A few studies have obtained rest-frame optical spectroscopy of intermediate-redshift SNe~Ia to test potential evolution with redshift. \citet{Coil00} presented peak brightness spectra of two SNe~Ia, SN~1999ff and SN~1999fv, at redshifts $z=0.46$ and $z=1.2$, respectively. Neither of these SNe displayed any properties that would suggest a deviation from the low-$z$ sample of SNe~Ia. \citet{Foley12:sdss} presented an analysis of SN~Ia discovered by the Sloan Digital Sky Survey-II (SDSS-II) SN Survey and followed up with Keck spectroscopy. This sample covered a redshift range of $0.11 \leq z \leq 0.37$. The SN~Ia spectra obtained were near peak brightness, covering near-UV to optical wavelengths. This study found that the Keck/SDSS SN~Ia sample exhibited a UV flux excess when compared with a representative low-$z$ sample. Several theoretical studies have suggested that progenitor metallicity, density structure, temperature, ionization, and dust all have the potential to impact observed UV-fluxes \citep{Hoflich98, Lentz00, Sauer08, Hachinger13, Mazzali14}. If the observed UV flux excess in the Keck/SDSS sample is the result of a metallicity effect, we would expect this trend to strengthen with redshift. At $z>2$, \textit{James Webb Space Telescope} (\textit{JWST}) has the necessary wavelength coverage to perform this test with a high-$z$ sample. 

Characterizing SN~Ia at $z>2$ is only possible with the \textit{JWST}. \textit{JWST} is capable of observing near-infrared (NIR) multi-band photometry down to $m_{\rm AB}^{}\sim 30$ mag and low-resolution (R$\sim30-300$) spectroscopy down to $m_{\rm AB}^{}\sim 29$ mag \citep{Decoursey24, siebert_discovery_2024, Pierel24, coulter_discovery_2025, Fox26}. This enables the construction of SN~Ia rest-frame optical-NIR light curves and spectroscopic classification from $z \sim 2-4$. 

Thus far, two non-lensed SNe~Ia have been observed with \textit{JWST} in the dark matter-dominated Universe at $z>2$ \citep{Pierel24, pierel_testing_2025}. The first, SN~2023adsy at $z=2.903$, was intrinsically red ($B - V \sim 0.8$ mag; but see \citealt{Vinko25}) and had a high \ion{Ca}{2} velocity ($\sim 19{,}000$ \kms). The second, SN~2023aeax at $z=2.15$, had a less extreme but peculiar blue color ($B - V \sim -0.3$ mag). Applying typical standardization techniques, together these SNe hint at a potential slope of measured distances with redshift (though in agreement with current evolving dark energy constraints, \citealt{pierel_testing_2025}), but a larger sample is needed to validate this result. 

While these SNe~Ia offer the first glimpse into the properties of the high-$z$ sample, their spectra were observed well-after peak brightness, making comparisons with low-$z$ samples more difficult. Furthermore, the spectra of these SNe also had considerable host contamination, which precluded making detailed measurements of SN spectral properties. 

In this work, we present \CosmosSNIa, a SN~Ia discovered in \textit{JWST} imaging as part of the COSMOS-3D Survey, a program which supplements the \textit{JWST} Cosmic Evolution Survey (COSMOS-Web; \citealt{casey_cosmos-web_2023}) with additional slitless spectroscopy and imaging. This survey obtained $\sim 100$ arcmin$^2$ of \textit{JWST}/NIRCam images, that serendipitously overlap with those from the Public Release IMaging for Extragalactic Research (PRIMER; \citealt{dunlop_primer_2021}) program. In contrast to a deep and narrow survey like the JWST Advanced Deep Extragalactic Survey (JADES; \citealt{Eisenstein23}), a wide-area JWST program like COSMOS is more likely to uncover younger explosions, and in particular, SNe Ia near peak \citep{Fox26}. Of the transients discovered in this data, we identified one SN~Ia candidate at $z\sim2$. We then activated follow-up observations using an existing \textit{JWST} target-of-opportunity program (ToO; PID 5324) to obtain a multi-band light curve and low-resolution classification spectrum. These observations yielded the first \textit{peak-brightness} classification spectrum of a high-$z$ SN~Ia, allowing for a novel detailed comparison to the low-$z$ sample.

In this paper, we begin \autoref{s:obs} by summarizing these observations. In Sections \ref{s:lc} - \ref{s:hd}, we measure a spectroscopic redshift, describe our light curve fitting procedure, place \CosmosSNIa on the Hubble diagram, and put this SN in the context of the other observed high-$z$ SNe~Ia. In Sections \ref{s:spec_comp} - \ref{s:vel}, we make use of {\tt kaepora} \citep{Siebert19} to provide a detailed comparison of our rest-frame near-UV to optical spectrum to the properties of low-$z$ SNe~Ia. We discuss our conclusions and considerations for future observations in \autoref{s:conc}. In this analysis, we assume a standard flat $\Lambda$CDM cosmology with $H_{0} = 70$ \kms $Mpc^{-1}$, $\Omega_{m} = 0.315$.

\section{Observations \& Data Reduction}\label{s:obs}

\begin{figure*}[htb!]
    \centering
    \includegraphics[width=5in]{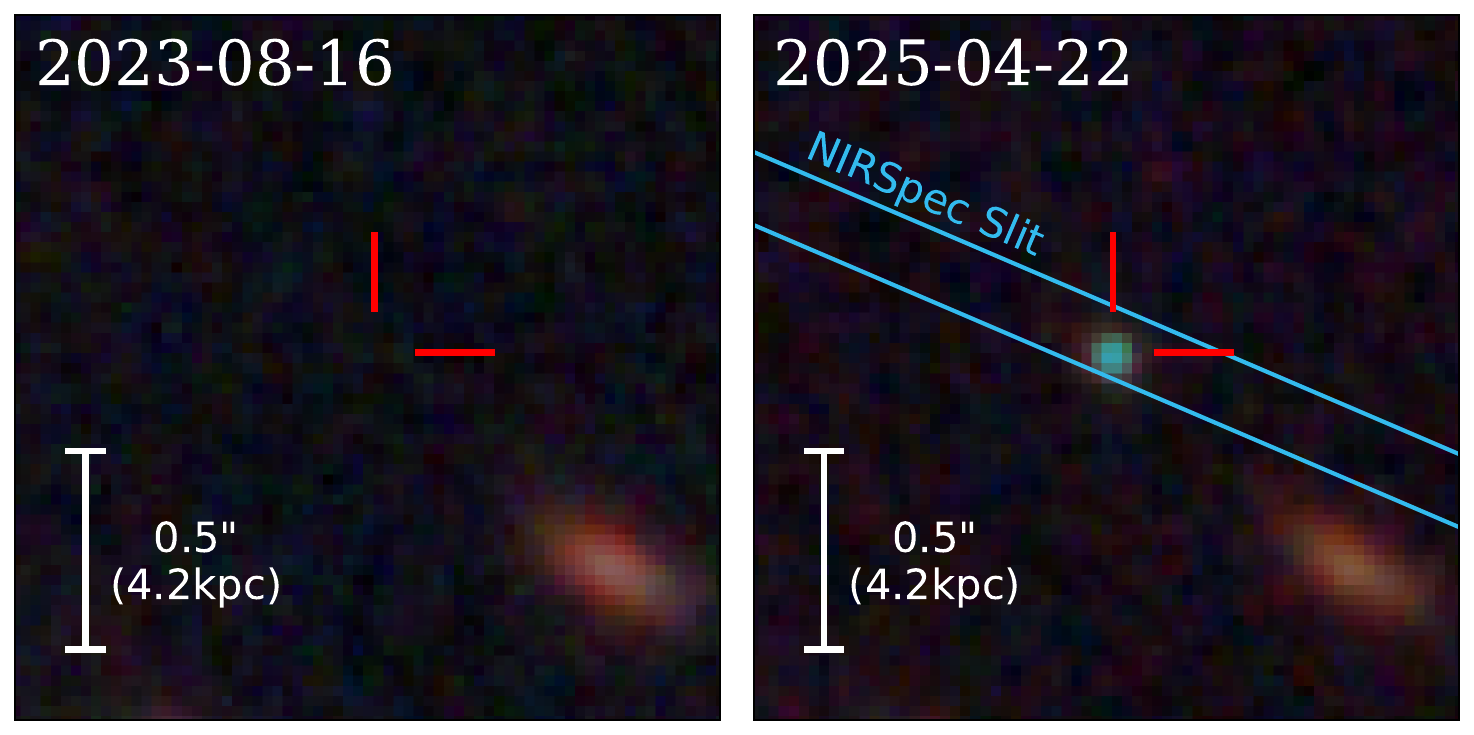}
    \caption{False color images of \CosmosSNIa and its host galaxy using F150W (Blue), F200W (Green), and F356W (Red). The left panel is the template image obtained from the PRIMER program (PID: 1837, \citealt{dunlop_primer_2021}) and the right panel is our discovery image from COSMOS-3D (PID: 5893). The SN is well-separated from its host galaxy at a projected distance of 8.1 kpc.}
    \label{fig:disc}
\end{figure*}

\CosmosSNIa was identified in COSMOS-3D imaging at R.A. $=10$h$00$m$27.8404$s and decl. $=+2$d$13$m$31.4406$s in three filters (F115W, F200W, and F356W) on April 22, 2025 via comparison to reference images from Aug 16, 2023 (\autoref{fig:disc}). The transient was offset from its host galaxy (COSMOS ID 7254, \citealt{Shuntov25}) by 0.969 arcseconds (a physical separation of 8.1 kpc) and modeling of the host galaxy SED yielded a photo-$z$ of $z = 2.0\pm0.1$, $\log(M_{*}/M_{\odot})$ = $8.6 \pm 0.1$, and $\log(\mathrm{sSFR})$ = $-8.5 \pm 0.6$. Preliminary light curve fits to the SN SED were consistent with a SN~Ia near peak brightness at this redshift.


\CosmosSNIa thus met the trigger criteria for our \textit{JWST} SN~Ia follow-up ToO program (PID 5324). This program allows for 10 total non-disruptive ToO activations to measure SN~Ia luminosity evolution with redshift. We activated this program to obtain two additional epochs of multi-band NIRCam photometry to characterize the light curve shape of \CosmosSNIa, and one epoch of NIRSpec low-resolution spectroscopy to provide a classification and spectroscopic redshift. In this section, we describe the data products and reduction methods for these observations.

\subsection{\textit{JWST} NIRCam} \label{s:obs_phot}

We follow the same methods for photometry on Level 3 (drizzled, I2D) \textit{JWST} images as \citet{pierel_discovery_2024,pierel_testing_2025}. Level 3 NIRCam images are the drizzled and resampled combination of Level 2 (CAL) NIRCam images, which are individual exposures that have been calibrated using the STScI JWST Pipeline\footnote{\url{https://github.com/spacetelescope/jwst}}
\citep{bushouse_jwst_2022},  
and have been bias-subtracted, dark-subtracted, and flat-fielded but not yet corrected for geometric distortion. 

We first align the individual NIRCam exposures containing \CosmosSNIa, from each program described above (see Table \ref{tab:phot}), to the drizzled template images from PRIMER, as PRIMER provides a temporal reference image in all filters where \CosmosSNIa was observed. We use the \textit{JWST}/\textit{HST} Alignment Tool \citep[{\tt JHAT};][]{rest_arminrestjhat_2023}\footnote{\url{https://jhat.readthedocs.io}}, which improves the relative default alignment from $\sim1$\,pixel to $\sim0.1$\,pixel between the epochs. Each CAL image is also corrected for 1/f noise using the default pipeline method.\footnote{\url{https://github.com/chriswillott/jwst}}

We produce aligned drizzled images with the \textit{JWST} pipeline \citep[v$1.19.2$;][]{bushouse_jwst_2022}, and obtain difference images in all filters using the High Order Transform of PSF and Template Subtraction \citep[{\tt HOTPANTS};][]{becker_hotpants_2015}\footnote{\url{https://github.com/acbecker/hotpants}} code \citep[with modifications implemented in the \texttt{photpipe} code;][]{rest_testing_2005}. We build a drizzed PSF using the \texttt{spike} package \citep{polzin_spike_2025}, which takes Level 2 PSF models from {\tt webbpsf}\footnote{\url{https://webbpsf.readthedocs.io}} that are temporally and spatially dependent and drizzles them together using the same pipeline implementation as the data. We then implement the \texttt{space\_phot} \citep{pierel_space-phot_2024}\footnote{\url{space-phot.readthedocs.io}} drizzled PSF fitting routine using $5\times5$ pixel cutouts, and fit to the observed \CosmosSNIa flux in all filters and epochs. These total fluxes, which are in units of MJy/sr, are converted to AB magnitudes using the native pixel scale of each image ($0.03\arcsec/$pix for SW, $0.06\arcsec/$pix for LW). Measured photometry is given in Table \ref{tab:phot}. A final source of photometric uncertainty is a systematic uncertainty on the zero-points, which is $\lesssim0.01$ mag for all filters and is therefore subdominant to the uncertainties derived here \citep{boyer_jwst_2022}. 


\begin{table}
    \centering
    \caption{\label{tab:phot} Photometry for \CosmosSNIa measured in Section \ref{s:obs_phot}.}
    
    \begin{tabular*}{\linewidth}{@{\extracolsep{\stretch{1}}}*{5}{c}}
\toprule
PID&Instrument&MJD&\multicolumn{1}{c}{Filter}&\multicolumn{1}{c}{$m_{\rm AB}^{}$}\\
\hline
$1837$&NIRCam&$60172$&F115W&$-$\\
$1837$&NIRCam&$60172$&F200W&$-$\\
$1837$&NIRCam&$60172$&F356W&$-$\\
\hline
$5893$&NIRCam&$60787$&F115W&$25.56\pm0.03$\\
$5893$&NIRCam&$60787$&F200W&$25.77\pm0.02$\\
$5893$&NIRCam&$60787$&F356W&$26.93\pm0.04$\\
\hline
$5324$&NIRCam&$60802$&F115W&$25.56\pm0.06$\\
$5324$&NIRCam&$60802$&F150W&$25.38\pm0.05$\\
$5324$&NIRCam&$60802$&F200W&$25.63\pm0.05$\\
$5324$&NIRCam&$60802$&F277W&$26.66\pm0.09$\\
$5324$&NIRCam&$60802$&F356W&$27.54\pm0.20$\\
$5324$&NIRCam&$60802$&F444W&$27.86\pm0.31$\\
\hline
$5324$&NIRCam&$60814$&F115W&$26.10\pm0.07$\\
$5324$&NIRCam&$60814$&F150W&$25.67\pm0.04$\\
$5324$&NIRCam&$60814$&F200W&$25.87\pm0.05$\\
$5324$&NIRCam&$60814$&F277W&$26.98\pm0.10$\\
$5324$&NIRCam&$60814$&F356W&$28.17\pm0.27$\\
$5324$&NIRCam&$60814$&F444W&$28.17\pm0.41$\\
\hline
\hline
    \end{tabular*}
\begin{flushleft}
\end{flushleft}
\vspace{1cm}
\end{table}

\subsection{\textit{JWST} NIRSpec} \label{s:obs_spec}
\CosmosSNIa was selected as one of the highest-priority targets for spectroscopic follow-up observations. We observed \CosmosSNIa on May 6, 2025 with \textit{JWST} NIRSpec in the Fixed Slits (FS) Spectroscopy mode \citep{Jakobsen2022, Birkmann2022, Rigby2022}. These observations used the S200A1 (0.200" wide $\times$ 3.300" long) slit with the prism grating and CLEAR filter. The total exposure time was 4011.945s. 

\begin{figure*}[htb!]
    \centering
    \includegraphics[width=6.4in]{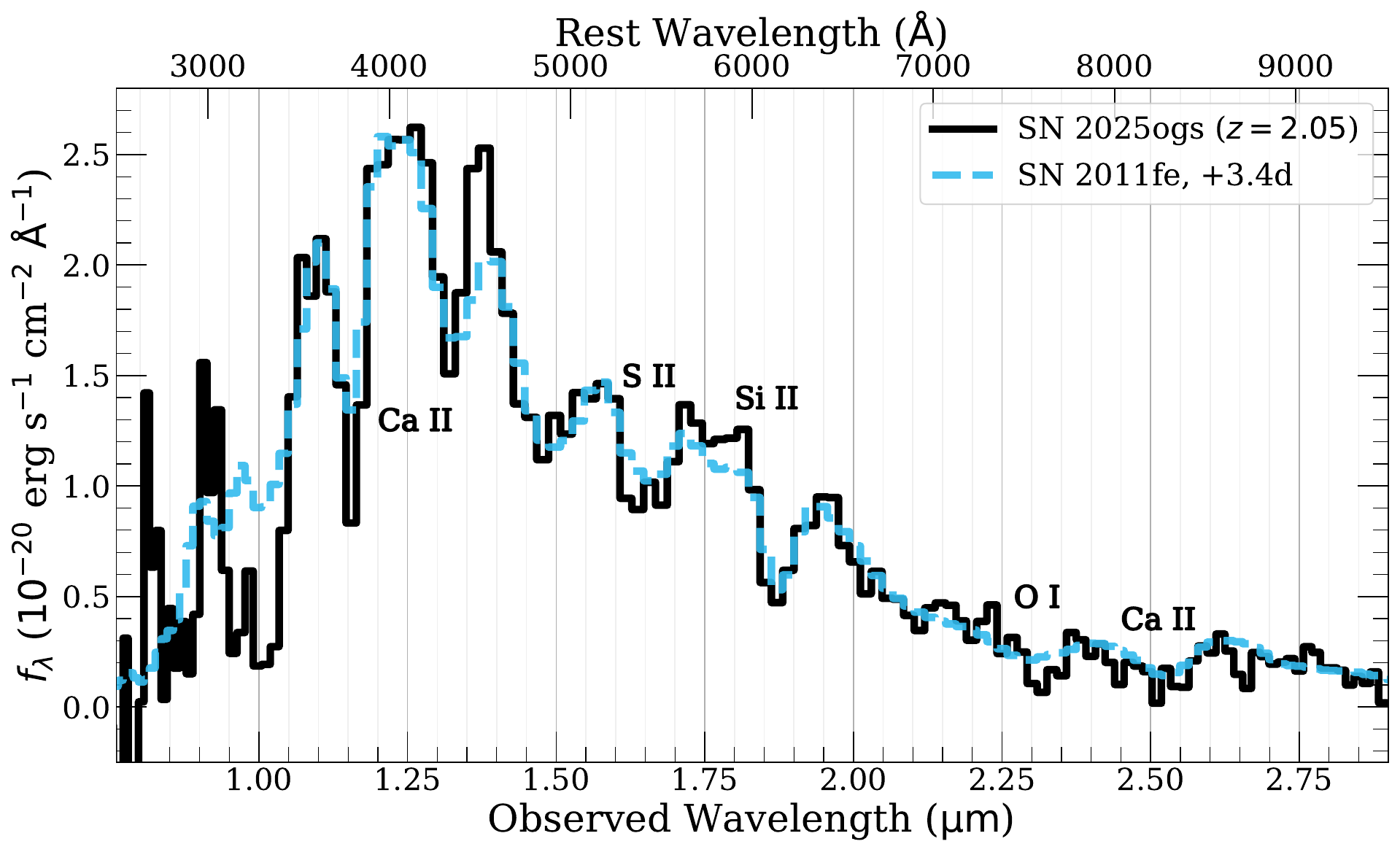}
    \caption{Comparison of our NIRSpec prism spectrum of \CosmosSNIa (black) to SN~2011fe (blue) at 3.4 days after peak brightness. The spectrum of SN~2011fe has been convolved with the prism dispersion function, rebinned, and scaled to match the observation. 
    \label{fig:2011fe}}
\end{figure*}

We reduced the \textit{JWST} data using the ``\textit{JWST}''\footnote{\url{https://github.com/spacetelescope/jwst}} pipeline \citep[version 1.19.1;][]{Bushouse_JWST_Calibration_Pipeline_2024} routines for bias and dark subtraction, background subtraction, flat-field correction, wavelength calibration, flux calibration, rectification, outlier detection, and resampling. The flux of the host galaxy was very faint relative to the SN (F200W $>29.5$ mag), therefore, we assumed there was negligible contamination. We performed an optimal-extraction of the SN using the default pipeline extraction parameters for point sources. These data are presented in \autoref{fig:2011fe}.

\section{Analysis \& Results}\label{s:anal}

\subsection{Spectroscopic Classification and Redshift}\label{s:spec_class}
We first perform spectroscopic classification and redshift measurement using the template-fitting code Next Generation SuperFit ({\tt NGSF}, \citealt{Goldwasser22})\footnote{\url{https://github.com/oyaron/NGSF}}. This code allows for the simultaneous fitting of the host galaxy light contamination and SN flux. In all subsequent fits, we restrict the fit the phase to $-3 < t < 5$ days (based on light curve fit from \autoref{s:lc}), and the extinction parameter to $0 < A_V < 2$ mag. 

Since we do not have a spectroscopic redshift from the host galaxy, we ran an initial fit over a coarse redshift grid ($1 < z < 3, \Delta z = 0.1$). The fit confirmed a classification of SN~Ia at $z\approx 2$. To obtain a more precise redshift, we restricted the fit to a finer and narrower redshift range ($2.0 < z < 2.2, \Delta z = 0.001$) resulting in a best match to SN~2011fe at $+2.8$ days and best-fit redshift of $z = 2.05$. The top five best-fit SNe~Ia have phases ranging  $-2.8 < t < 2.8$ days, and redshifts ranging $2.046 < z < 2.065$. All best-fitting SN templates indicate that the SN contributes $>99\%$ of the total flux, consistent with the relative large separation of \CosmosSNIa from its host galaxy. In \autoref{fig:2011fe} we show a comparison of \CosmosSNIa to SN~2011fe at a similar phase of $+3.4$ days which enables a comparison of the flux in the NUV. 

\begin{deluxetable}{ccc}
    \tablecaption{Best-fit parameters from \CosmosSNIa spectroscopy \label{tab:specfit}}
    \tablehead{\colhead{Parameter} & \colhead{Kaepora ($1.1 - 2.7 \mu m$)} & \colhead{Kaepora ($1.3 - 2.2 \mu m$)}}
    \tablecolumns{3}
        \startdata
        $z$ & $2.06 \pm 0.01$ & $2.049 \pm 0.005$\\
        $t$ & $+1.8 \pm 1.9$ & $+1.0 \pm 2.2$\\
        $\Delta m_{15} (B)$ & $1.18 \pm 0.15$ & $+1.22 \pm 0.17$\\
        \enddata
    \footnotesize{$^*$ Uncertainties derived from the standard deviation of best-fit templates.}\\
\end{deluxetable}

To refine our redshift estimate, we constructed a redshift fitting routine using optical spectra of normal-SN~Ia from {\tt kaepora}, an open-source database for SN~Ia spectra \citep{Siebert19}. {\tt kaepora} is a relational database and suite of analysis tools for SN~Ia spectra. This database contains a large number of homogenized SN~Ia spectra (deredshifted, interpolated, and corrected from Milky Way and host galaxy reddening) and associated light curve metadata. Our procedure was as follows: first, we selected spectra from {\tt kaepora} from SNe~Ia with $-3 < t < 3$ days and $1.0 < \Delta m_{15} (B) < 1.75$ with wavelength ranges covering $3500 - 9500$ in the rest frame (resulting in 66 spectra of 43 SNe~Ia). We then convolved these spectra with the NIRSpec prism dispersion function for $z = 2.046$ (assuming minimal differences over our narrow redshift fitting range). For each spectrum we then determined the best-fit redshift which minimized $\chi^2$ over observed wavelengths of $1.1 - 2.7\mu m$. We obtain $z = 2.06 \pm 0.01$ from the distribution of best fitting spectra with $\chi^2/\nu < 2$ (for SNe with multiple spectra in the sample, only the highest ranked is used in this average). We note that SN~2011fe ($z=2.049$) is ranked second using this method. 

$z = 2.06$ is slightly higher than the valued determined by {\tt NGSF}, however, if we exclude \ion{Ca}{2} features, ($1.3 - 2.2\mu m$) we obtain $z = 2.049 \pm 0.005$. This discrepancy results primarily from a difference between the \ion{Si}{2} and \ion{Ca}{2} ejecta velocities relative to low-$z$ SNe~Ia which is discussed in detail in \autoref{s:vel}. We summarize our redshift fitting results in \autoref{tab:specfit}.

The {\tt kaepora} redshift fit using the larger wavelength range results in a \ion{Si}{2} ejecta velocity of $-12{,}200$ \kms which would make \CosmosSNIa a HV SN~Ia. Given the good match to SN~2011fe (which is not a HV SN~Ia) using multiple methods, we favor the lower value solution. Therefore, we adopt $z = 2.05 \pm 0.01$ for \CosmosSNIa for the remainder of this work. However, we discuss the implications of assuming a slightly larger redshift in \autoref{s:vel} (see \autoref{fig:spec_meas}).

\subsection{Light Curve}\label{s:lc}
\begin{figure}[htb!]
    \centering
    \includegraphics[width=3.2in]{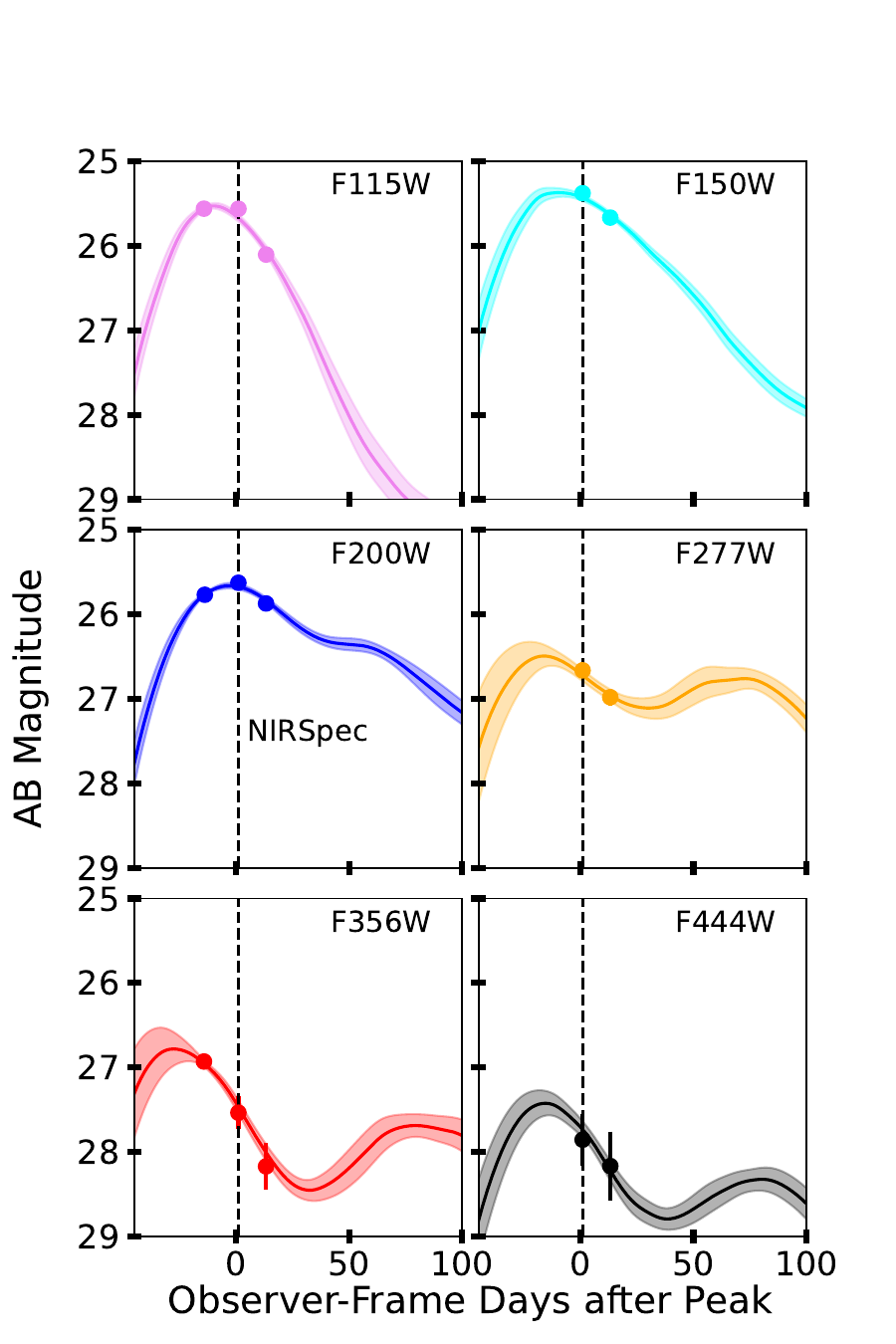}
    \caption{Best fit {\tt BayeSN} model to each of the observed filters and epochs for \CosmosSNIa. The measured photometry is shown as circles with error, and the shaded region represents the {\tt BayeSN} model$+$ fitting uncertainty. The epoch of our NIRSpec observation is noted by the dashed-black vertical line.
    \label{fig:lc}}
\end{figure}
\begin{figure}[htb!]
    \centering
    \includegraphics[width=2.6in]{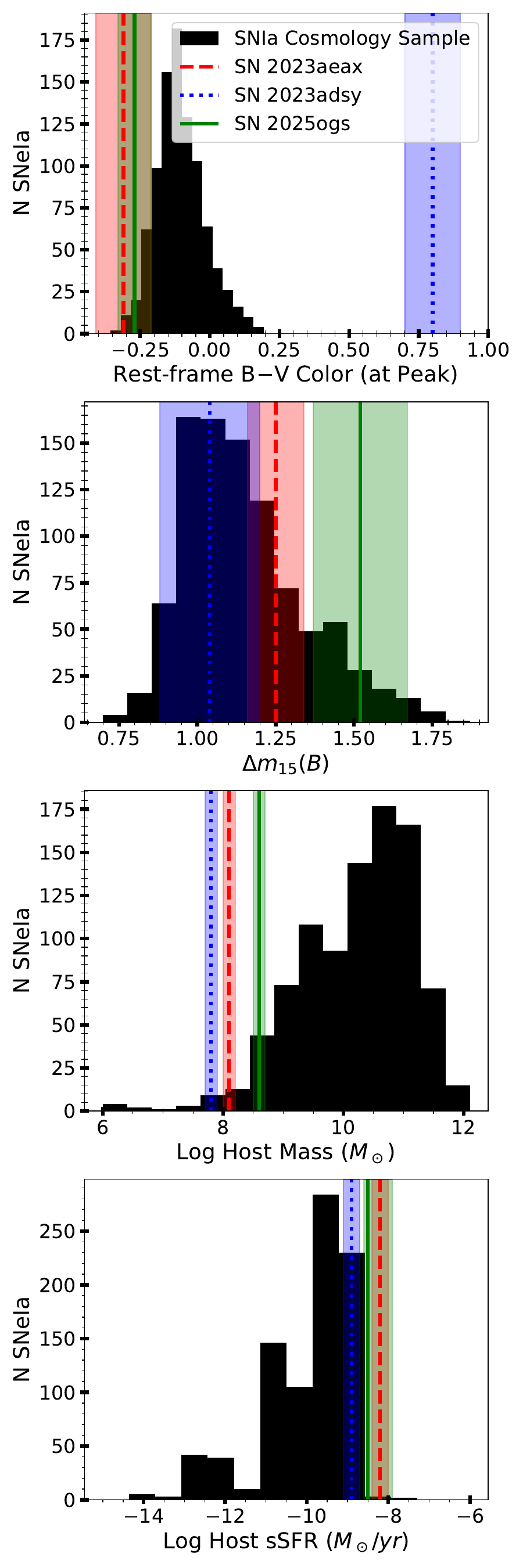}
    \caption{The distributions of rest-frame B$-$V color at peak B-band brightness, $\Delta m_{15}(B)$, and host galaxy properties measured for the cosmological sample of SNe\,Ia presented in \citet{brout_pantheon_2022}. The vertical lines with shaded $1\sigma$ uncertainties are \CosmosSNIa (green solid) and the other spectroscopic SNe\,Ia, namely SN\,2023adsy (blue dotted) and SN\,2023aeax (red dashed).
    \label{fig:params}}
\end{figure}
We fit the observed photometry (Table \ref{tab:phot}) with the {\tt BayeSN} SN\,Ia SED model \citep{mandel_hierarchical_2022,ward_relative_2023,grayling_scalable_2024}, which provides robust rest-frame UV-NIR SN\,Ia light curves fits. The version of {\tt BayeSN} we use is presented by \citet{ward_relative_2023}, and is trained to cover rest-frame phases as late as $50$ days after peak-brightness. The model has been shown to produce excellent fits of sparse high-$z$ SN\,Ia light curves \citep[e.g.,][]{pierel_testing_2025}.

\begin{figure*}[htb!]
    \centering
    \includegraphics[width=5.5in]{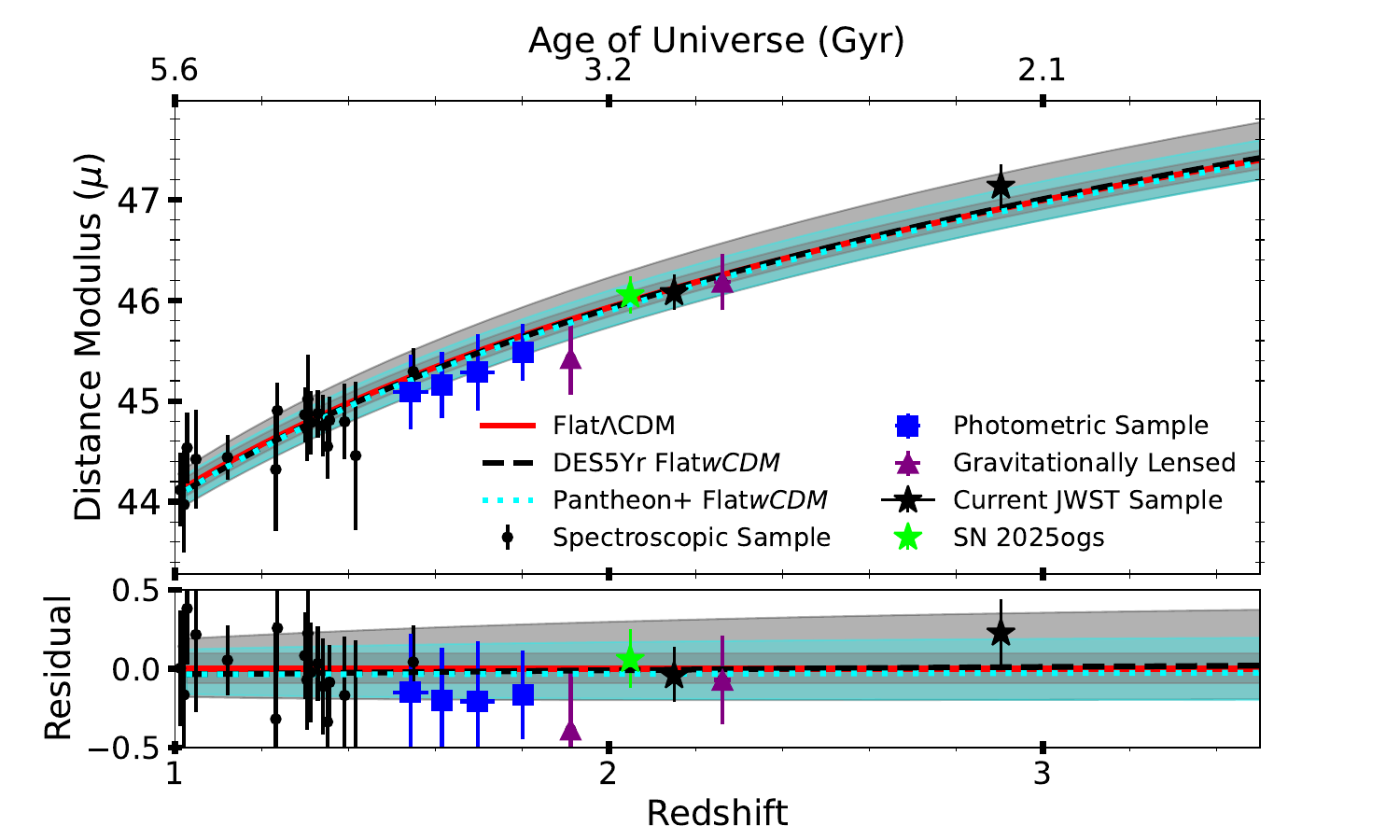}
    \caption{Luminosity distance measurements from SNe\,Ia at $z>1$ including data from \citet{brout_pantheon_2022}. Black points (with errors) are SNe\,Ia with spectroscopic classifications, while blue squares (with error) are SNe\,Ia with photometric classifications. The two strongly lensed SNe\,Ia with distance measurements are shown as purple triangles. \CosmosSNIa is shown as a green star, and Flat$\Lambda$CDM ($H_0=70$ km s$^{-1}$ Mpc$^{-1}$) is shown as a solid red line. The black stars are the distance modulii for the other $z>2$ spectroscopically confirmed SNe\,Ia with \textit{JWST}. The best-fit Flat$wCDM$ cosmological constraints from \citet{brout_pantheon_2022} (black-dashed) and \citet{des_collaboration_dark_2024} (cyan-dotted) are also shown for comparison.
    \label{fig:dist}}
\end{figure*}
In addition to the base SED template, which is based on the model from \citet{hsiao_k_2007}, {\tt BayeSN} includes a rest-frame host-galaxy dust component (parameterized by the $V$-band extinction $A_V$ and ratio of total to selective extinction $R_V$) and a light curve shape parameter $\theta$. We add a $E(B - V)=0.01$\,mag Galactic extinction correction to the model based on the dust maps of \citet{schlafly_measuring_2011} and using the extinction curve from \citet{fitzpatrick_correcting_1999}. The {\tt BayeSN} ``$\epsilon$-surface'' captures all additional model/intrinsic scatter and intrinsic color variation. Finally, {\tt BayeSN} directly infers the luminosity distances as part of the Bayesian inference process. The best-fit model is shown with the observed photometry in Figure \ref{fig:lc}, and the retrieved {\tt BayeSN} parameters from the fit are shown in Table \ref{tab:lcfit}. The value of the light curve parameter $\theta$ corresponds to a $\Delta m_{15}(B)$ of $\sim1.55\pm0.15$ mag (the total decline in $B$-band brightness 15 days after maximum-light), which is within the normal low-$z$ SN\,Ia population but somewhat fast-declining (Figure \ref{fig:params}). The rest-frame $B-V$ color at peak brightness for \CosmosSNIa is $-0.27\pm0.06$~mag, which is on the blue edge of the normal low-$z$ SN population but in agreement with a previous $z=2.15$ SN\,Ia discovery \citep[see Figure \ref{fig:params};][]{pierel_testing_2025} and within traditional low-$z$ color cosmology cuts \citep[$-0.4\lesssim B-V\lesssim0.4$ mag, or using the SALT ``$c$'' parameter $-0.3<c<0.3$;][]{brout_pantheon_2022}.

\begin{deluxetable}{ccc}
    \tablecaption{Best-fit light curve parameters and properties for \CosmosSNIa \label{tab:lcfit}}
    \tablehead{\colhead{Parameter} & \colhead{Bounds} & \colhead{SN~Ia}}
    \tablecolumns{3}
        \startdata
        $z$&Fixed&$z=\snz$ \\
        $t_{pk}$&[$60750,60850$]& $60796.38_{-2.43}^{+1.73}$\\
        $\mu$&[0,$100$]& $46.05_{-0.18}^{+0.19}$\\
        $A_V$&[$0,2$]& $<0.08$\\
        $\theta$&[$-3,3$]&$0.07_{-0.59}^{+0.44}$\\
        \enddata
\end{deluxetable}

\subsection{Adding \CosmosSNIa to the Hubble Diagram}\label{s:hd}
As noted in the previous section, {\tt BayeSN} infers the luminosity distance for a SN\,Ia directly as a result of its Bayesian model fitting step. Following previous work in the high-$z$ regime \citep{pierel_discovery_2024,pierel_testing_2025} we do not explicitly apply a bias correction \citep[e.g.,][]{kessler_correcting_2017} or correction for the host-galaxy mass step. We apply half of the host mass step (for a low-mass galaxy) from \citet{brout_pantheon_2022} (who found $\sim0.054$mag) and add a systematic error of half the host mass step in quadrature. As was the case in \citet{pierel_testing_2025}, \CosmosSNIa is several magnitudes brighter than the survey limiting magnitude and has normal light curve parameters, meaning that a bias correction will be negligible. 

The final luminosity distance measurement is $\mu=46.05_{-0.18}^{+0.19}$\,mag, while the flat $\Lambda$CDM prediction at $z=\snz$ (with $H_0=70$ km s$^{-1}$ Mpc$^{-1}$) is $\mu=46.02$\,mag, a $<1\sigma$ difference (\autoref{fig:dist}). The uncertainty on $\mu$ includes the fitted model uncertainties, errors peculiar velocity (which are negligible here), and an additional $0.055$\,mag uncertainty from weak gravitational lensing \citep{jonsson_constraining_2010}. Similarly, we ignored the additional uncertainty from the redshift, which we tested during the fitting stage and found to be sub-dominant. The distance measurement from \CosmosSNIa is in excellent agreement with SN\,2023aeax at $z=2.15$, suggesting minimal evolution in SN\,Ia standardizability at $z\sim2$. To further quantify this, we fit a flat $\Lambda$CDM model using only the \textit{JWST} sample. Assuming a linear dependence of the residuals with $(1+z)$ or age, we find that these observations constrain luminosity evolution to $0.29 \pm 0.26$ mag/$(1+z)$ or $0.26 \pm 0.23$ mag/Gyr, respectively.

\subsection{Spectroscopic Comparisons to low-$z$ SNe~Ia}\label{s:spec_comp}

\begin{figure*}[htb!]
    \centering
    \includegraphics[width=6.4in]{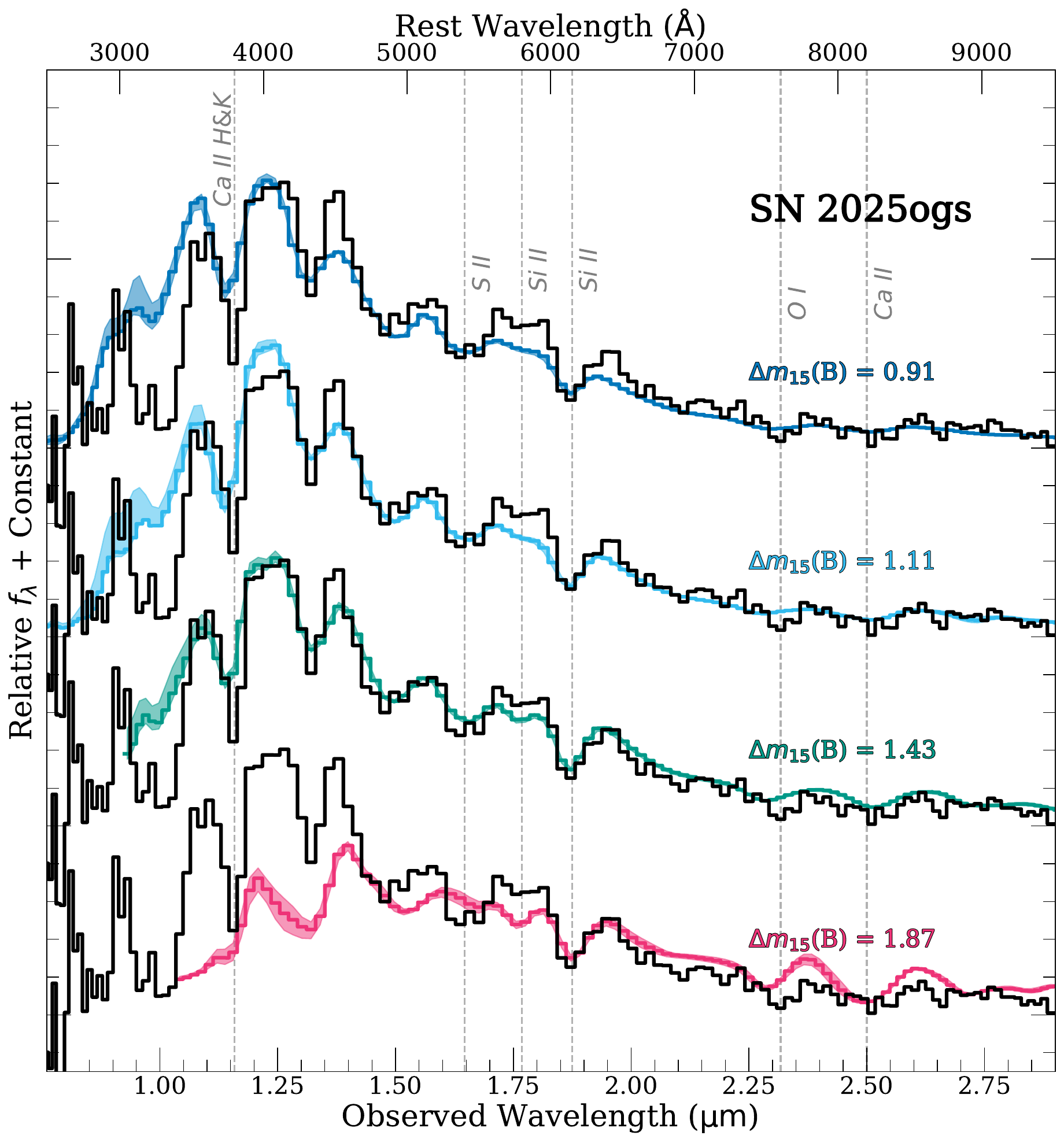}
    \caption{Comparison of \CosmosSNIa (black) to maximum brightness ($-3 < t < 3$ days) composite spectra generated from {\tt kaepora} \citep{Siebert19} for a variety of light curve shapes (colored curves). The shaded regions are the $1\sigma$ uncertainties estimated via bootstrap resampling. The mean $\Delta m_{15}(B)$ of the composite spectra range from 0.91 to 1.87. Overall the continuum and features are most similar to the $\Delta m_{15}(B) = 1.43$ composite spectrum (consistent with the light curve). The largest differences appear to be in the NUV ($<3500 \AA$) and in the velocity of some spectral features (notably, \ion{Ca}{2} H\&K, \ion{O}{1} $\lambda 7774$, and the \ion{Ca}{2} NIR triplet).
    \label{fig:kpora_max}}
\end{figure*}

While the low resolution of the NIRSpec prism typically complicates classification at high-$z$ \citep{siebert_discovery_2024, Pierel24, coulter_discovery_2025}, the spectrum of \CosmosSNIa unambiguously matches the features of SN~Ia observed near peak brightness. In \autoref{fig:2011fe}, we compare the spectrum of \CosmosSNIa (black curve) to an optical (+3.4 day) spectrum of SN~2011fe that has been redshifted to $z= \snz$ and convolved with the dispersion function of the NIRSpec prism (blue curve). We observe distinct absorption features from \ion{Ca}{2} H\&K , \ion{S}{2} $\lambda \lambda$ 6716, 6731, \ion{Si}{2} $\lambda$ 5972, 6355, \ion{O}{1} $\lambda $ 7774, and the \ion{Ca}{2} NIR triplet, which are all characteristic and well-studied in SNe~Ia at this phase. Unlike other high-$z$ SNe observed with \textit{JWST} \citep{siebert_discovery_2024, coulter_discovery_2025}, our spectrum does not show any narrow emission lines from the host galaxy due to its fairly large separation. Furthermore, the strengths of the absorption features are qualitatively similar to SN~2011fe, indicating that there is likely little contamination from the host galaxy continuum. 


To test the similarity of \CosmosSNIa with the low-$z$ sample of SNe~Ia, we make use of {\tt kaepora} composite spectra. Using the same procedures discussed by \citet{Siebert19, siebert_possible_2020} we generate peak-brightness composite spectra for a variety of SN~Ia light curve shapes ($\Delta m_{15}$ ($B$) = 0.91, 1.11, 1.43, and 1.87 mag, generated from 103, 193, 139 and 36, total spectra, respectively). As done for the spectrum of SN~2011fe, we perform the additional necessary steps of redshifting these composite spectra to $z=\snz$ and convolving them with the NIRSpec prism dispersion function. These composite spectra are shown in \autoref{fig:kpora_max} (colored curves) and compared with our spectrum of \CosmosSNIa (black curve). These composite spectra serve as a high-level visualization of how the average spectral properties of SN~Ia populations vary with light curve shape, and are useful for qualitative comparison.

Overall, aside from the fastest declining subset (pink curve), the peak brightness composite spectra are remarkably similar to \CosmosSNIa. The $0.91 < \Delta m_{15}$ ($B$) $< 1.43$ composite spectra match the \ion{Ca}{2} H\&K, \ion{S}{2} $\lambda \lambda$ 6716, 6731, \ion{Si}{2} $\lambda$ 5972, 6355, \ion{O}{1} $\lambda $ 7774, and the \ion{Ca}{2} NIR triplet very well yet there are some subtle differences. The largest deviation occurs in the rest-frame near-UV where the composite spectra all over-predict the relative flux of \CosmosSNIa. \CosmosSNIa is broadly consistent the composite spectra with mean $\Delta m_{15}$ ($B$) = 1.11 and 1.43. These composite spectra show similar colors, and ratio of the peaks blueward and redward of \ion{Ca}{2} H\&K. We note that, at this redshift, the \ion{Ca}{2} H\&K and \ion{Ca}{2} NIR triplet absorption velocities of \CosmosSNIa appear less blueshifted than in each of the composite spectra. This is explored further in \autoref{s:vel}.

\begin{figure*}[htb!]
    \centering
    \includegraphics[width=2.9in]{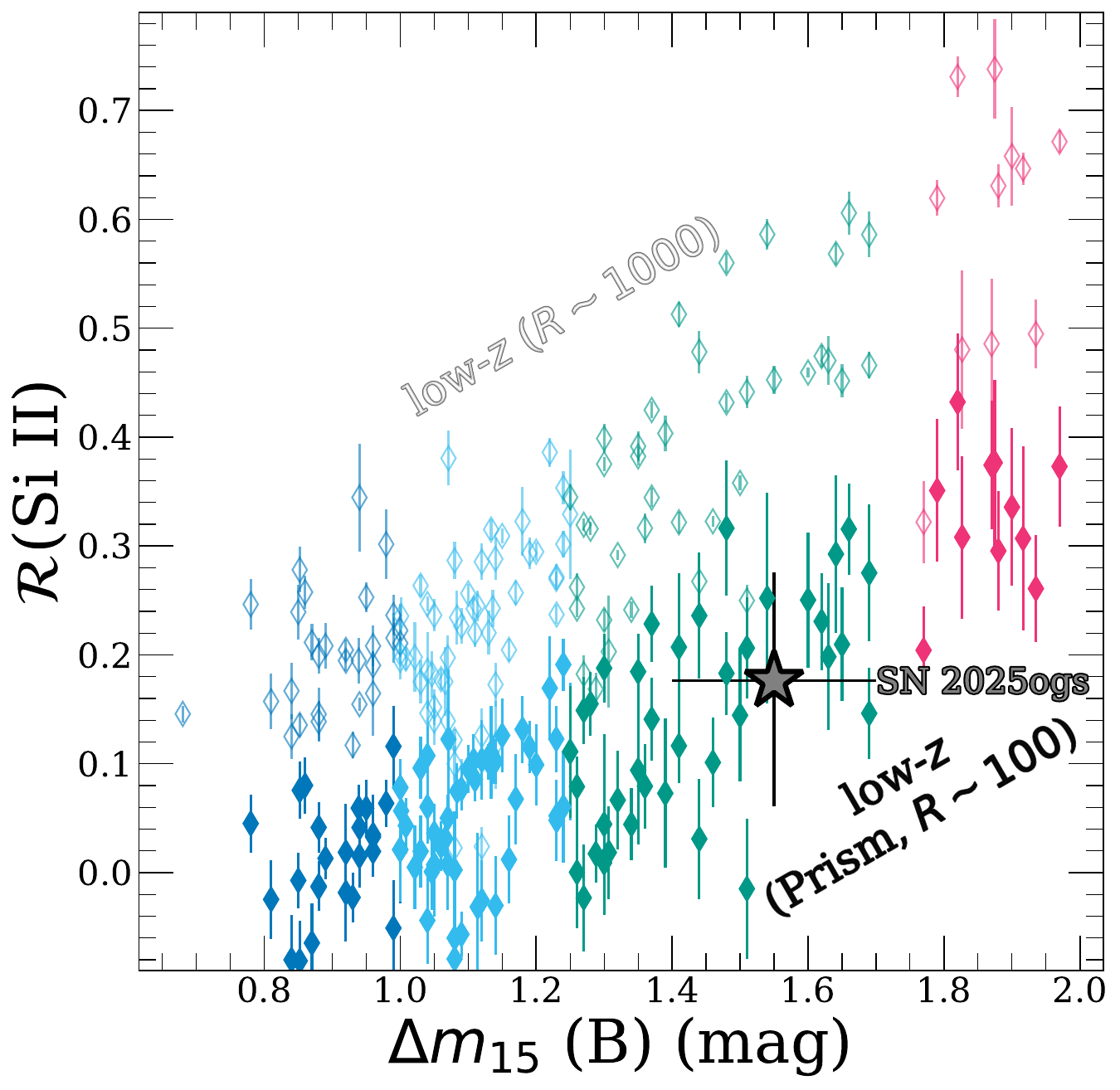}
    \includegraphics[width=3.51in]{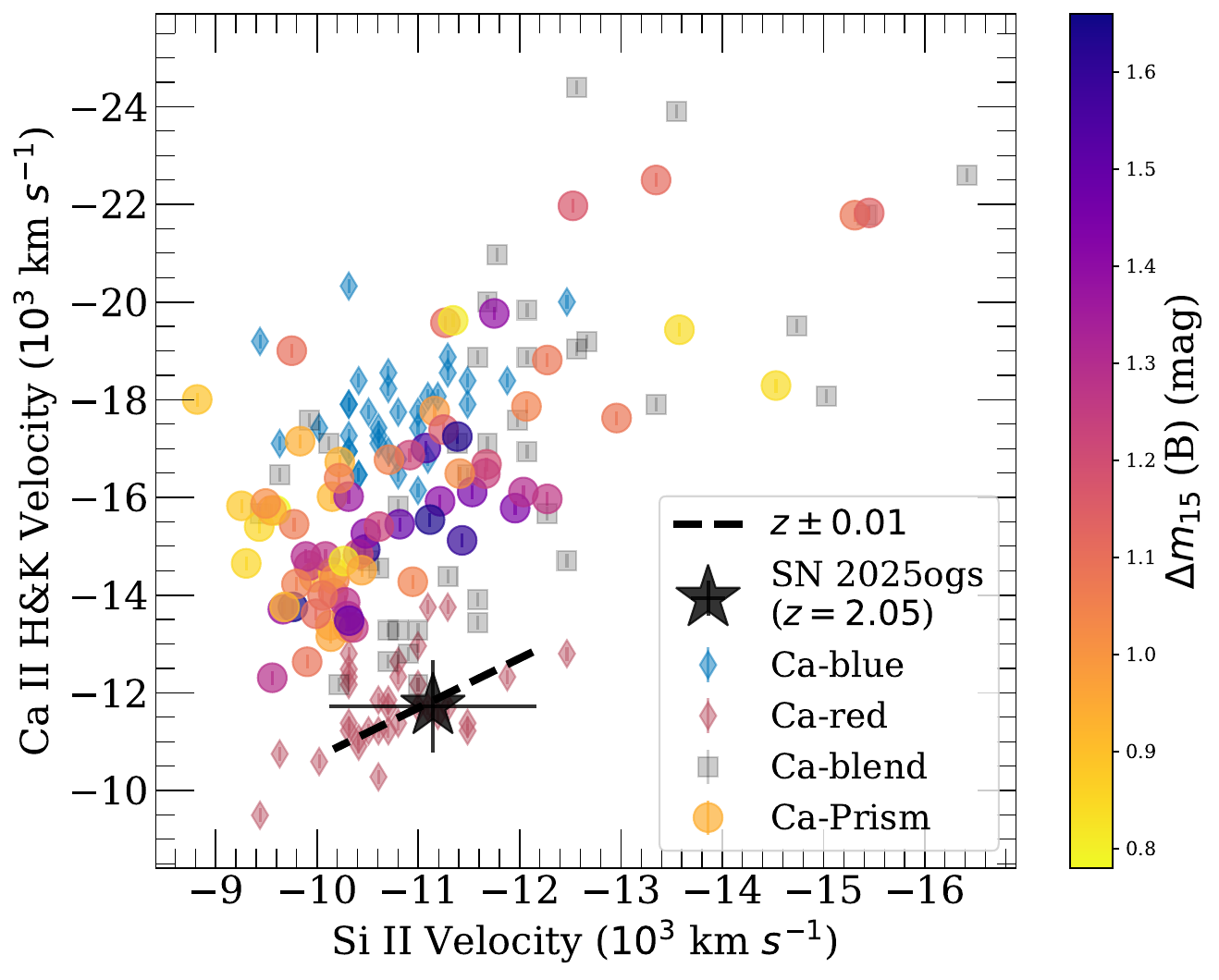}
    \caption{\textit{left}: Comparison of $\mathcal{R}$(\ion{Si}{2}) (the ratio of the absorption depth of \ion{Si}{2} $\lambda 5972$ to \ion{Si}{2}$\lambda 6355$; \citealt{Nugent95}) of \CosmosSNIa (black star) to the low-$z$ sample (colored points) vs $\Delta m_{15}(B)$. The color  of the points reference the corresponding composite spectrum (from \autoref{fig:kpora_max}) to which these SNe contribute. The open points correspond to  measurements derived from the individual spectra at their original resolution. The closed points result from convolving these individual spectra with the NIRSpec prism dispersion function and re-measuring $\mathcal{R}$(\ion{Si}{2}). \textit{right}: Comparison of the \ion{Si}{2} and \ion{Ca}{2} H\&K ejecta velocities of \CosmosSNIa (black star) to the low-$z$ sample (colored points) using the techniques illustrated in \autoref{fig:vel_meas}. Light-colored diamonds correspond measurements from the low-$z$ sample using spectra at their original resolution. The red points result from the photospheric component of \ion{Ca}{2} H\&K, while the blue points are from the high-velocity component. The gray squares (Ca-blend) represent velocities measured from the minimum of the \ion{Ca}{2} H\&K feature when two separate components are not clearly visible. The colored-circles (color indicating the value of $\Delta m_{15}(B)$) result from measuring the minimum of \ion{Ca}{2} H\&K after convolving the low-$z$ spectra with the NIRSpec prism dispersion function. The black-dashed line indicates how the conservative uncertainty on redshift ($z\pm0.01$) would impact the ejecta velocity measurements of \CosmosSNIa. While the \ion{Si}{2} velocity is consistent with the low-$z$ sample, the \ion{Ca}{2} H\&K velocity is significantly lower.}
    \label{fig:spec_meas}
\end{figure*}

The relative depth of these \ion{Si}{2} (the ``Si Ratio'', $\mathcal{R}$(\ion{Si}{2})) features have a well-studied relationship with light-curve shape \citep{Nugent95,Phillips99}. \CosmosSNIa has a moderately-fast decline rate measured from the light curve ($\Delta m_{15}$ ($B$) = 1.55, \autoref{s:lc}). To see if this is consistent with the spectrum, we compare $\mathcal{R}$(\ion{Si}{2}) of \CosmosSNIa to the low-$z$ spectra that contribute to the {\tt kaepora} composite spectra in \autoref{fig:spec_meas} (left). The color of the diamond points correspond to the composite spectra (from \autoref{fig:kpora_max}) to which these SNe contribute. The open points represent measurements made from the spectra at their original resolution, and the closed points represent measurements made after convolving these spectra with the NIRSpec prism dispersion function. The gray star shows our \CosmosSNIa measurement of $\mathcal{R}$(\ion{Si}{2}) from the spectrum ($0.17^{+0.12} _{-0.09}$) and $\Delta m_{15}$ ($B$) from the light curve ($1.55\pm0.15$ mag). 

The change of resolution results in a clear bias in the value of $\mathcal{R}$(\ion{Si}{2}). The average offset of the unfilled/filled points is $\Delta \mathcal{R}$(\ion{Si}{2})$=0.21$. Additionally, measurements made using the prism resolution are mostly consistent with $\mathcal{R}$(\ion{Si}{2}) $= 0$ below $\Delta m_{15}$ ($B$) $\sim 1.1$, consistent with the flattening of \ion{Si}{2} $\lambda$ 5972 seen in the blue composite spectra from \autoref{fig:kpora_max}. Despite this offset, the prism-$\mathcal{R}$(\ion{Si}{2}) values are still strongly correlated with $\Delta m_{15}(B)$.
We find that $\mathcal{R}$(\ion{Si}{2}) of \CosmosSNIa is consistent with what is expected for its light curve shape. However, given the relatively large uncertainty, and the best matching spectra from redshift fitting, $\Delta m_{15}(B) = 1.22 \pm 0.17$ (see \autoref{tab:specfit}), we cannot definitively distinguish between a proto-typical SN~Ia (like SN~2011fe) or something more transitional based on the spectrum alone. Given the properties of the light curve fit, and closer similarity of the $\Delta m_{15}(B) = 1.43$ composite spectrum (particularly in the NUV), we suggest a slight preference for a faster decline rate. 

\subsection{Ejecta Velocity}\label{s:vel}

Given that we do not have a spectroscopic redshift from the host galaxy, the redshift uncertainty ($\sigma_{z} = 0.01$) will limit our ability to characterize the ejecta velocities of \CosmosSNIa. From the comparisons in \autoref{fig:kpora_max} it is clear that there may be interesting differences in relative velocities that are worth investigating. For example, while the minima of the \ion{Si}{2} features are well-matched between \CosmosSNIa and the composite spectra, the \ion{Ca}{2} and \ion{O}{1} features of  \CosmosSNIa appear to be at systematically lower velocities. 

To understand these differences, we inspect the properties of the population of SNe~Ia that contribute to the composite spectra. First, we select SN~Ia spectra from {\tt kaepora} with phase between $-3$ and $+3$ days and $\Delta m_{15}$ ($B$) $< 1.75$ mag. Then, events with multiple spectra in this range are combined to create a single representative max-light SED. As previously done, we then redshift these spectra to the redshift of \CosmosSNIa and convolve them with the dispersion function of the NIRSpec prism. Finally, we create a smoothed spectrum using the inverse-variance Gaussian smoothing algorithm from \citet{Blondin06}, interpolate to a finer wavelength grid, and measure the wavelength where flux is at a minimum. Notably, at this resolution, the individual velocity components of \ion{Ca}{2} H\&K are not resolved, resulting in an absorption minimum that is representative of a blend of both components. For a detailed illustration of this process see \autoref{fig:vel_meas} in \autoref{a:appendix}. We measure a \ion{Si}{2} $\lambda 6355$ ejecta velocity of $-11,100\pm1000$ \kms and (blended) \ion{Ca}{2} H\&K ejecta velocity of $-11,700\pm1100$ \kms via this method\footnote{The \ion{Ca}{2} H\&K minimum wavelength has been converted to velocity assuming a rest wavelength of 3945\AA\ (the gf-weighted average of the 3933 and 3969\AA\ features)}. Velocity uncertainties have been estimated via a Monte Carlo simulation that incorporates the uncertainty on flux and the uncertainty in optimal choice of smoothing parameter. The effects of redshift uncertainty are shown in the right panel of \autoref{fig:spec_meas} (dashed black line)

In \autoref{fig:spec_meas} (right), we show a comparison of \ion{Si}{2} $\lambda$ 6355 and \ion{Ca}{2} H\&K (prism flux minimum) ejecta velocities (colored circles) of the individual SN~Ia spectra to those measured in \CosmosSNIa (black star). The red and blue diamonds correspond to the ejecta velocities measured from the photospheric- and high-velocity components of \ion{Ca}{2} H\&K from the original spectra with higher resolution. 

First, we find that the \ion{Si}{2} $\lambda$6355 ejecta velocities of the original and prism-convolved spectra are similar, while the multiple components of \ion{Ca}{2} H\&K pull our prism-convolved ejecta velocity measurements toward higher values. Interestingly, we do not observe a similar offset in velocity in the spectrum of \CosmosSNIa: the prism-convolved ejecta velocity of \ion{Ca}{2} H\&K is significantly lower than one would expect based on the population of low-$z$ SNe Ia. Specifically, low-$z$ SNe~Ia with consistent \ion{Si}{2} velocities, have a mean \ion{Ca}{2} H\&K velocity of $-15,500\pm1600$ \kms corresponding to a $2\sigma$ difference from \CosmosSNIa ($-11,700\pm1100$ \kms). Only two SNe~Ia in the low-$z$ sample have prism-convolved \ion{Ca}{2} H\&K velocities consistent with \CosmosSNIa, but these SNe have relatively low \ion{Si}{2} velocities. Additionally, as described previously, the minimum of the \ion{Ca}{2} NIR triplet also appears less blueshifted than those in the composite spectra, pointing toward a lower average velocity. Using a similar procedure to obtain a representative ejecta velocity from the minimum of this feature, we find a value of $-10,900\pm2000$ \kms which is below (but consistent with) the low-$z$ sample mean of $-13,100\pm2500$ \kms.\footnote{Assumes a rest wavelength of 8579\AA\ for the \ion{Ca}{2} NIR triplet} These results could potentially indicate that \CosmosSNIa does not have a high-velocity \ion{Ca}{2} component.

As described in \autoref{s:spec_class}, the exact wavelength range over which one uses to fit the redshift does impact these velocity measurements. The black-dashed line depicts the impact of redshift uncertainty on both the \ion{Si}{2} and \ion{Ca}{2} velocities. We note, though, that even with the conservative estimate of the redshift uncertainty, the ejecta velocity as measured by the \ion{Ca}{2} H\&K feature in \CosmosSNIa remains anomalous compared with the low-$z$ sample.

\section{Discussion \& Conclusions}\label{s:conc}

We have presented \textit{JWST} photometric and spectroscopic follow-up observations of \CosmosSNIa. These observations have facilitated the unambiguous classification of this transient as a normal-SN~Ia near peak-brightness at $z = \snz$. Its multi-band light curve indicates a blue color ($B - V = -0.27\pm 0.06$ mag) similar to SN 2023aeax at $z = 2.15$ ($B - V = -0.3$ mag), and moderate decline rate $\Delta m_{15} (B) = 1.55 \pm 0.15$ mag. Through standardization of its light curve, we find that it is within $1\sigma$ of standard $\Lambda$CDM in agreement with other $z>2$ SNe~Ia \citep{Pierel24, pierel_testing_2025}. The existing \textit{JWST} SN~Ia sample constrains evolution of standardized luminosities in the dark matter-dominated Universe to $0.29 \pm 0.26$ mag/$(1+z)$ ($0.26 \pm 0.23$ mag/Gyr). While the data show no evidence for SN~Ia evolution with only moderate constraints, this represents the first test of SN~Ia evolution at $z > 2$, where a wide range of dark energy models (i.e., $w$CDM with $-1.5 < w < -0.5$) only predict distance modulus changes of $<$0.06~mag. A larger \textit{JWST} sample is still needed to distinguish between these effects.

The NIRSpec prism spectrum of \CosmosSNIa has offered a unique opportunity to compare its high-S/N spectral features to the low-$z$ SN~Ia sample. We find remarkable similarity to SN~2011fe, a quintessential normal-SN~Ia, at 3.4 days after peak brightness. We have produced composite spectra from the low-$z$ sample that simulate the resolution of the NIRSpec prism. Through these comparisons we both corroborate the measured light curve parameters and estimate the bias on spectral measurements observed with this instrumentation. 

While the \ion{Si}{2} velocity of \CosmosSNIa is uncertain ($-11,100\pm1000$ \kms), it is consistent with a low-velocity SN~Ia which is in agreement with its best-matched SNe (e.g., SN~2011fe) and its light curve properties. Despite this agreement, there are interesting differences between our measured \ion{Si}{2} and \ion{Ca}{2} and those measured from the low-$z$ sample that cannot be explained via redshift uncertainty. Future samples, should investigate whether high-$z$ SNe~Ia have significantly different \ion{Ca}{2} ejecta properties. 

The observations presented in this paper offer a new benchmark for comparison of both photometric and spectroscopic evolution of SN~Ia properties with redshift. We have developed methods to compare the spectroscopic properties of high-$z$ SNe~Ia to their low-$z$ counterparts. This analysis shows that there is promise in the prospect of characterizing the diversity of SN~Ia at high-$z$ with \textit{JWST}. With a modest exposure time ($\sim 4000$s) one can obtain S/N $\sim 10$ in many key features for a peak brightness SN~Ia at $z = 2$. With slightly longer exposure times one could constrain important line diagnostics like ejecta velocity and pseudo-equivalent width. Additional light curve epochs could significantly help to constrain the light curve shape which is needed for interpreting the spectroscopy. These observations can establish the foundational subclassification needed for higher-dimensional representations that can be helpful in identifying outliers (e.g., \citealt{Burrow20,Bose25}) and for the detailed modeling efforts that ultimately link spectral variations to physical differences in SN Ia progenitors. While detailed characterization like this at high-$z$ is only possible with \textit{JWST}, the methods developed for interpretation of the data are of critical importance for \textit{Roman}. 

\textit{Roman}'s spectroscopy is more limited in sensitivity than \textit{JWST}, however, its prism is similar in design and a significant fraction of the survey is devoted to spectroscopy. Understanding the biases associated with measurements of spectral features from low-resolution data will be valuable for understanding the evolution of SN~Ia properties out to $z < 1$ from \textit{Roman}. 

Taken together, these results demonstrate that normal SNe~Ia exist in the dark matter-dominated Universe. The distance inferred for \CosmosSNIa is in agreement with $\Lambda$CDM suggesting that cosmological constraints from \textit{Roman}'s high-precision observations will remain robust. Although the spectroscopic properties of \CosmosSNIa are broadly consistent with the low-redshift population, the subtle differences we identify, particularly in the \ion{Ca}{2} and \ion{Si}{2} ejecta velocities, underscore the need for larger, well-characterized samples to determine whether such variations reflect genuine redshift evolution or simply the diversity already present in the local Universe. 

\begin{center}
    \textbf{Acknowledgements}
\end{center}

This work is based on observations made with the NASA/ESA/CSA \textit{JWST} as part of programs 1837, 5893, and 5324. The STScI TSST group acknowledges partial support from  JWST-GO-06541, JWST-GO-06585, and JWST-GO-05324. This letter is based on observations with the NASA/ESA Hubble Space Telescope and James Webb Space Telescope obtained from the Mikulski Archive for Space Telescopes at STScI at doi:\dataset[10.17909/6n92-8508]{https://archive.stsci.edu/doi/resolve/resolve.html?doi=10.17909/6n92-8508}. We thank the \textit{JWST}/HST scheduling teams and instrument experts at STScI for extraordinary effort in getting the observations used here scheduled quickly. 

MRS is supported by an STScI Postdoctoral Fellowship. JP is supported by NASA through a Einstein Fellowship grant No. HF2-51541.001 awarded by the Space Telescope Science Institute (STScI), which is operated by the Association of Universities for Research in Astronomy, Inc., for NASA, under contract NAS5-26555. DOJ acknowledges support from NSF grants AST-2407632, AST-2429450, and AST-2510993, NASA grant 80NSSC24M0023, and HST/JWST grants HST-GO-17128.028 and JWST-GO-05324.031, awarded by the Space Telescope Science Institute (STScI), which is operated by the Association of Universities for Research in Astronomy, Inc., for NASA, under contract NAS5-26555. DCL acknowledges support from STScI grant JWST-GO-05324.002-A, under which part of this research was carried out.

\facilities{\textit{JWST} (NIRCam/NIRSpec)}

\software{astropy \citep{astropy,astropy2,astropy3}}

\newpage

\appendix 

\section{} \label{a:appendix}

Example simulated velocity measurements of SN~2011fe using the spectral properties of the NIRSpec prism. The left and right panels are zoomed in on \ion{Ca}{2} H\&K and \ion{Si}{2} $\lambda$ 5972,6355, respectively. The thin colored curves are SN~2011fe at its original resolution and the thick colored curves are the same SN~2011fe spectra convolved with the NIRSpec prism resolution function at $z=\snz$. The yellow circles mark are our measurements of the minium flux from which we derive our ejecta velocities. The black curves show \CosmosSNIa for comparison. It is important to note that at the prism resolution, the photospheric- and high-velocity components of \ion{Ca}{2} H\&K are not individually resolved. This means that the ejecta velocity that we derive from the convolved spectrum is really a combination of both features that can be compared with a similar measurement from the NIRSpec prism spectrum of \CosmosSNIa. 
\begin{figure*}[htb!]
    \centering
    \includegraphics[width=3.2in]{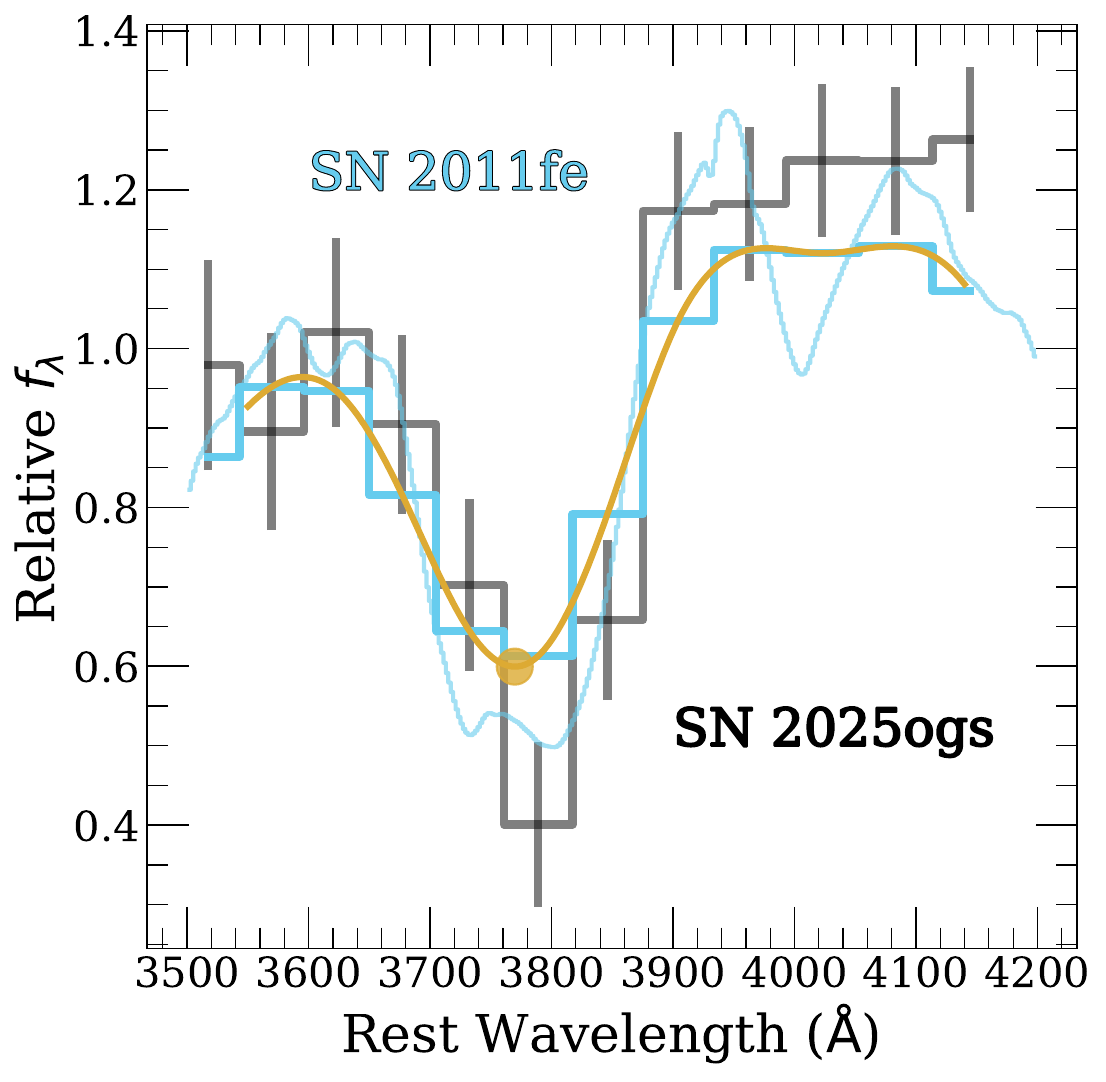}
    \includegraphics[width=3.2in]{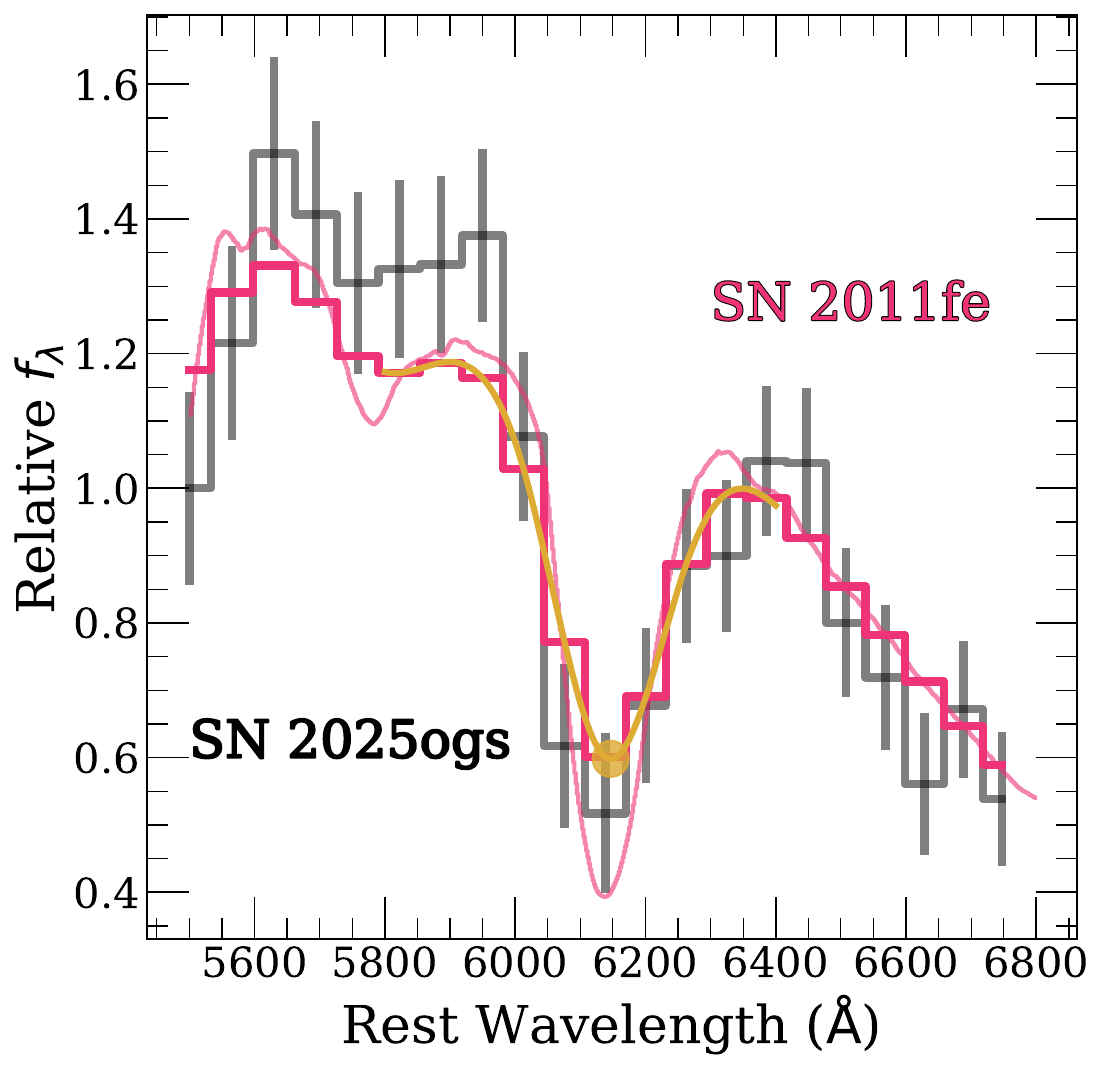}
    \caption{\textit{left}: Example measurement of a \textit{JWST} prism-convolved \ion{Ca}{2} H\&K ejecta velocity that can be inferred from NIRSpec prism spectra at this redshift. The thin-blue colored curve is SN~2011fe at peak brightness in its original resolution. The thick-blue colored curve is the same spectrum, but convolved with the NIRSpec prism dispersion function. We measure the velocity from the minimum of the yellow curve (smoothed and interpolated from the thick-blue curve). At this resolution, the low-velocity photospheric component and high-velocity component cannot be independently measured. We show \CosmosSNIa (black curve) for comparison, the minimum of \ion{Ca}{2} H\&K is less blueshifted. 
    \textit{right}: Same as the left panel, but for \ion{Si}{2}. Here, there is only one velocity component contributing to the measurement. 
    \label{fig:vel_meas}}
\end{figure*}


\newpage
\bibliography{astro_refs,pierel_refs}

@ARTICLE{jha25,
       author = {{Observations Time Allocation Committee}, Roman and {Community Survey Definition Committees}, Core},
        title = "{Roman Observations Time Allocation Committee: Final Report and Recommendations}",
      journal = {arXiv e-prints},
         year = 2025,
        month = may,
          eid = {arXiv:2505.10574},
        pages = {arXiv:2505.10574},
          doi = {10.48550/arXiv.2505.10574},
archivePrefix = {arXiv},
       eprint = {2505.10574},
 primaryClass = {astro-ph.IM},
       adsurl = {https://ui.adsabs.harvard.edu/abs/2025arXiv250510574O}
}

@ARTICLE{LSST_Sci_Col09,
       author = {{LSST Science Collaboration} and {Abell}, Paul A. and {Allison}, Julius and {Anderson}, Scott F. and {Andrew}, John R. and {Angel}, J. Roger P. and {Armus}, Lee and {Arnett}, David and {Asztalos}, S.~J. and {Axelrod}, Tim S. and {Bailey}, Stephen and {Ballantyne}, D.~R. and {Bankert}, Justin R. and {Barkhouse}, Wayne A. and {Barr}, Jeffrey D. and {Barrientos}, L. Felipe and {Barth}, Aaron J. and {Bartlett}, James G. and {Becker}, Andrew C. and {Becla}, Jacek and {Beers}, Timothy C. and {Bernstein}, Joseph P. and {Biswas}, Rahul and {Blanton}, Michael R. and {Bloom}, Joshua S. and {Bochanski}, John J. and {Boeshaar}, Pat and {Borne}, Kirk D. and {Bradac}, Marusa and {Brandt}, W.~N. and {Bridge}, Carrie R. and {Brown}, Michael E. and {Brunner}, Robert J. and {Bullock}, James S. and {Burgasser}, Adam J. and {Burge}, James H. and {Burke}, David L. and {Cargile}, Phillip A. and {Chandrasekharan}, Srinivasan and {Chartas}, George and {Chesley}, Steven R. and {Chu}, You-Hua and {Cinabro}, David and {Claire}, Mark W. and {Claver}, Charles F. and {Clowe}, Douglas and {Connolly}, A.~J. and {Cook}, Kem H. and {Cooke}, Jeff and {Cooray}, Asantha and {Covey}, Kevin R. and {Culliton}, Christopher S. and {de Jong}, Roelof and {de Vries}, Willem H. and {Debattista}, Victor P. and {Delgado}, Francisco and {Dell'Antonio}, Ian P. and {Dhital}, Saurav and {Di Stefano}, Rosanne and {Dickinson}, Mark and {Dilday}, Benjamin and {Djorgovski}, S.~G. and {Dobler}, Gregory and {Donalek}, Ciro and {Dubois-Felsmann}, Gregory and {Durech}, Josef and {Eliasdottir}, Ardis and {Eracleous}, Michael and {Eyer}, Laurent and {Falco}, Emilio E. and {Fan}, Xiaohui and {Fassnacht}, Christopher D. and {Ferguson}, Harry C. and {Fernandez}, Yanga R. and {Fields}, Brian D. and {Finkbeiner}, Douglas and {Figueroa}, Eduardo E. and {Fox}, Derek B. and {Francke}, Harold and {Frank}, James S. and {Frieman}, Josh and {Fromenteau}, Sebastien and {Furqan}, Muhammad and {Galaz}, Gaspar and {Gal-Yam}, A. and {Garnavich}, Peter and {Gawiser}, Eric and {Geary}, John and {Gee}, Perry and {Gibson}, Robert R. and {Gilmore}, Kirk and {Grace}, Emily A. and {Green}, Richard F. and {Gressler}, William J. and {Grillmair}, Carl J. and {Habib}, Salman and {Haggerty}, J.~S. and {Hamuy}, Mario and {Harris}, Alan W. and {Hawley}, Suzanne L. and {Heavens}, Alan F. and {Hebb}, Leslie and {Henry}, Todd J. and {Hileman}, Edward and {Hilton}, Eric J. and {Hoadley}, Keri and {Holberg}, J.~B. and {Holman}, Matt J. and {Howell}, Steve B. and {Infante}, Leopoldo and {Ivezic}, Zeljko and {Jacoby}, Suzanne H. and {Jain}, Bhuvnesh and {R} and {Jedicke} and {Jee}, M. James and {Garrett Jernigan}, J. and {Jha}, Saurabh W. and {Johnston}, Kathryn V. and {Jones}, R. Lynne and {Juric}, Mario and {Kaasalainen}, Mikko and {Styliani} and {Kafka} and {Kahn}, Steven M. and {Kaib}, Nathan A. and {Kalirai}, Jason and {Kantor}, Jeff and {Kasliwal}, Mansi M. and {Keeton}, Charles R. and {Kessler}, Richard and {Knezevic}, Zoran and {Kowalski}, Adam and {Krabbendam}, Victor L. and {Krughoff}, K. Simon and {Kulkarni}, Shrinivas and {Kuhlman}, Stephen and {Lacy}, Mark and {Lepine}, Sebastien and {Liang}, Ming and {Lien}, Amy and {Lira}, Paulina and {Long}, Knox S. and {Lorenz}, Suzanne and {Lotz}, Jennifer M. and {Lupton}, R.~H. and {Lutz}, Julie and {Macri}, Lucas M. and {Mahabal}, Ashish A. and {Mandelbaum}, Rachel and {Marshall}, Phil and {May}, Morgan and {McGehee}, Peregrine M. and {Meadows}, Brian T. and {Meert}, Alan and {Milani}, Andrea and {Miller}, Christopher J. and {Miller}, Michelle and {Mills}, David and {Minniti}, Dante and {Monet}, David and {Mukadam}, Anjum S. and {Nakar}, Ehud and {Neill}, Douglas R. and {Newman}, Jeffrey A. and {Nikolaev}, Sergei and {Nordby}, Martin and {O'Connor}, Paul and {Oguri}, Masamune and {Oliver}, John and {Olivier}, Scot S. and {Olsen}, Julia K. and {Olsen}, Knut and {Olszewski}, Edward W. and {Oluseyi}, Hakeem and {Padilla}, Nelson D. and {Parker}, Alex and {Pepper}, Joshua and {Peterson}, John R. and {Petry}, Catherine and {Pinto}, Philip A. and {Pizagno}, James L. and {Popescu}, Bogdan and {Prsa}, Andrej and {Radcka}, Veljko and {Raddick}, M. Jordan and {Rasmussen}, Andrew and {Rau}, Arne and {Rho}, Jeonghee and {Rhoads}, James E. and {Richards}, Gordon T. and {Ridgway}, Stephen T. and {Robertson}, Brant E. and {Roskar}, Rok and {Saha}, Abhijit and {Sarajedini}, Ata and {Scannapieco}, Evan and {Schalk}, Terry and {Schindler}, Rafe and {Schmidt}, Samuel},
        title = "{LSST Science Book, Version 2.0}",
      journal = {arXiv e-prints},
         year = 2009,
        month = dec,
          eid = {arXiv:0912.0201},
        pages = {arXiv:0912.0201},
          doi = {10.48550/arXiv.0912.0201},
archivePrefix = {arXiv},
       eprint = {0912.0201},
 primaryClass = {astro-ph.IM},
       adsurl = {https://ui.adsabs.harvard.edu/abs/2009arXiv0912.0201L}
}

@ARTICLE{LSST_Sci_Col17,
       author = {{LSST Science Collaboration} and {Marshall}, Phil and {Anguita}, Timo and {Bianco}, Federica B. and {Bellm}, Eric C. and {Brandt}, Niel and {Clarkson}, Will and {Connolly}, Andy and {Gawiser}, Eric and {Ivezic}, Zeljko and {Jones}, Lynne and {Lochner}, Michelle and {Lund}, Michael B. and {Mahabal}, Ashish and {Nidever}, David and {Olsen}, Knut and {Ridgway}, Stephen and {Rhodes}, Jason and {Shemmer}, Ohad and {Trilling}, David and {Vivas}, Kathy and {Walkowicz}, Lucianne and {Willman}, Beth and {Yoachim}, Peter and {Anderson}, Scott and {Antilogus}, Pierre and {Angus}, Ruth and {Arcavi}, Iair and {Awan}, Humna and {Biswas}, Rahul and {Bell}, Keaton J. and {Bennett}, David and {Britt}, Chris and {Buzasi}, Derek and {Casetti-Dinescu}, Dana I. and {Chomiuk}, Laura and {Claver}, Chuck and {Cook}, Kem and {Davenport}, James and {Debattista}, Victor and {Digel}, Seth and {Doctor}, Zoheyr and {Firth}, R.~E. and {Foley}, Ryan and {Fong}, Wen-fai and {Galbany}, Lluis and {Giampapa}, Mark and {Gizis}, John E. and {Graham}, Melissa L. and {Grillmair}, Carl and {Gris}, Phillipe and {Haiman}, Zoltan and {Hartigan}, Patrick and {Hawley}, Suzanne and {Hlozek}, Renee and {Jha}, Saurabh W. and {Johns-Krull}, C. and {Kanbur}, Shashi and {Kalogera}, Vassiliki and {Kashyap}, Vinay and {Kasliwal}, Vishal and {Kessler}, Richard and {Kim}, Alex and {Kurczynski}, Peter and {Lahav}, Ofer and {Liu}, Michael C. and {Malz}, Alex and {Margutti}, Raffaella and {Matheson}, Tom and {McEwen}, Jason D. and {McGehee}, Peregrine and {Meibom}, Soren and {Meyers}, Josh and {Monet}, Dave and {Neilsen}, Eric and {Newman}, Jeffrey and {O'Dowd}, Matt and {Peiris}, Hiranya V. and {Penny}, Matthew T. and {Peters}, Christina and {Poleski}, Radoslaw and {Ponder}, Kara and {Richards}, Gordon and {Rho}, Jeonghee and {Rubin}, David and {Schmidt}, Samuel and {Schuhmann}, Robert L. and {Shporer}, Avi and {Slater}, Colin and {Smith}, Nathan and {Soares-Santos}, Marcelles and {Stassun}, Keivan and {Strader}, Jay and {Strauss}, Michael and {Street}, Rachel and {Stubbs}, Christopher and {Sullivan}, Mark and {Szkody}, Paula and {Trimble}, Virginia and {Tyson}, Tony and {de Val-Borro}, Miguel and {Valenti}, Stefano and {Wagoner}, Robert and {Wood-Vasey}, W. Michael and {Zauderer}, Bevin Ashley},
        title = "{Science-Driven Optimization of the LSST Observing Strategy}",
      journal = {arXiv e-prints},
         year = 2017,
        month = aug,
          eid = {arXiv:1708.04058},
        pages = {arXiv:1708.04058},
          doi = {10.48550/arXiv.1708.04058},
archivePrefix = {arXiv},
       eprint = {1708.04058},
 primaryClass = {astro-ph.IM},
       adsurl = {https://ui.adsabs.harvard.edu/abs/2017arXiv170804058L}
}

@ARTICLE{astropy,
       author = {{Astropy Collaboration} and {Robitaille}, Thomas P. and
         {Tollerud}, Erik J. and {Greenfield}, Perry and {Droettboom}, Michael and
         {Bray}, Erik and {Aldcroft}, Tom and {Davis}, Matt and
         {Ginsburg}, Adam and {Price-Whelan}, Adrian M. and
         {Kerzendorf}, Wolfgang E. and {Conley}, Alexander and {Crighton}, Neil and
         {Barbary}, Kyle and {Muna}, Demitri and {Ferguson}, Henry and
         {Grollier}, Fr{\'e}d{\'e}ric and {Parikh}, Madhura M. and
         {Nair}, Prasanth H. and {Unther}, Hans M. and {Deil}, Christoph and
         {Woillez}, Julien and {Conseil}, Simon and {Kramer}, Roban and
         {Turner}, James E.~H. and {Singer}, Leo and {Fox}, Ryan and
         {Weaver}, Benjamin A. and {Zabalza}, Victor and {Edwards}, Zachary I. and
         {Azalee Bostroem}, K. and {Burke}, D.~J. and {Casey}, Andrew R. and
         {Crawford}, Steven M. and {Dencheva}, Nadia and {Ely}, Justin and
         {Jenness}, Tim and {Labrie}, Kathleen and {Lim}, Pey Lian and
         {Pierfederici}, Francesco and {Pontzen}, Andrew and {Ptak}, Andy and
         {Refsdal}, Brian and {Servillat}, Mathieu and {Streicher}, Ole},
        title = "{Astropy: A community Python package for astronomy}",
      journal = {\aap},
         year = 2013,
        month = oct,
       volume = {558},
          eid = {A33},
        pages = {A33},
          doi = {10.1051/0004-6361/201322068},
archivePrefix = {arXiv},
       eprint = {1307.6212},
 primaryClass = {astro-ph.IM},
       adsurl = {https://ui.adsabs.harvard.edu/abs/2013A&A...558A..33A}
}

@ARTICLE{astropy2,
       author = {{Astropy Collaboration} and {Price-Whelan}, A.~M. and {Sip{\H{o}}cz}, B.~M. and {G{\"u}nther}, H.~M. and {Lim}, P.~L. and {Crawford}, S.~M. and {Conseil}, S. and {Shupe}, D.~L. and {Craig}, M.~W. and {Dencheva}, N. and {Ginsburg}, A. and {VanderPlas}, J.~T. and {Bradley}, L.~D. and {P{\'e}rez-Su{\'a}rez}, D. and {de Val-Borro}, M. and {Aldcroft}, T.~L. and {Cruz}, K.~L. and {Robitaille}, T.~P. and {Tollerud}, E.~J. and {Ardelean}, C. and {Babej}, T. and {Bach}, Y.~P. and {Bachetti}, M. and {Bakanov}, A.~V. and {Bamford}, S.~P. and {Barentsen}, G. and {Barmby}, P. and {Baumbach}, A. and {Berry}, K.~L. and {Biscani}, F. and {Boquien}, M. and {Bostroem}, K.~A. and {Bouma}, L.~G. and {Brammer}, G.~B. and {Bray}, E.~M. and {Breytenbach}, H. and {Buddelmeijer}, H. and {Burke}, D.~J. and {Calderone}, G. and {Cano Rodr{\'\i}guez}, J.~L. and {Cara}, M. and {Cardoso}, J.~V.~M. and {Cheedella}, S. and {Copin}, Y. and {Corrales}, L. and {Crichton}, D. and {D'Avella}, D. and {Deil}, C. and {Depagne}, {\'E}. and {Dietrich}, J.~P. and {Donath}, A. and {Droettboom}, M. and {Earl}, N. and {Erben}, T. and {Fabbro}, S. and {Ferreira}, L.~A. and {Finethy}, T. and {Fox}, R.~T. and {Garrison}, L.~H. and {Gibbons}, S.~L.~J. and {Goldstein}, D.~A. and {Gommers}, R. and {Greco}, J.~P. and {Greenfield}, P. and {Groener}, A.~M. and {Grollier}, F. and {Hagen}, A. and {Hirst}, P. and {Homeier}, D. and {Horton}, A.~J. and {Hosseinzadeh}, G. and {Hu}, L. and {Hunkeler}, J.~S. and {Ivezi{\'c}}, {\v{Z}}. and {Jain}, A. and {Jenness}, T. and {Kanarek}, G. and {Kendrew}, S. and {Kern}, N.~S. and {Kerzendorf}, W.~E. and {Khvalko}, A. and {King}, J. and {Kirkby}, D. and {Kulkarni}, A.~M. and {Kumar}, A. and {Lee}, A. and {Lenz}, D. and {Littlefair}, S.~P. and {Ma}, Z. and {Macleod}, D.~M. and {Mastropietro}, M. and {McCully}, C. and {Montagnac}, S. and {Morris}, B.~M. and {Mueller}, M. and {Mumford}, S.~J. and {Muna}, D. and {Murphy}, N.~A. and {Nelson}, S. and {Nguyen}, G.~H. and {Ninan}, J.~P. and {N{\"o}the}, M. and {Ogaz}, S. and {Oh}, S. and {Parejko}, J.~K. and {Parley}, N. and {Pascual}, S. and {Patil}, R. and {Patil}, A.~A. and {Plunkett}, A.~L. and {Prochaska}, J.~X. and {Rastogi}, T. and {Reddy Janga}, V. and {Sabater}, J. and {Sakurikar}, P. and {Seifert}, M. and {Sherbert}, L.~E. and {Sherwood-Taylor}, H. and {Shih}, A.~Y. and {Sick}, J. and {Silbiger}, M.~T. and {Singanamalla}, S. and {Singer}, L.~P. and {Sladen}, P.~H. and {Sooley}, K.~A. and {Sornarajah}, S. and {Streicher}, O. and {Teuben}, P. and {Thomas}, S.~W. and {Tremblay}, G.~R. and {Turner}, J.~E.~H. and {Terr{\'o}n}, V. and {van Kerkwijk}, M.~H. and {de la Vega}, A. and {Watkins}, L.~L. and {Weaver}, B.~A. and {Whitmore}, J.~B. and {Woillez}, J. and {Zabalza}, V. and {Astropy Contributors}},
        title = "{The Astropy Project: Building an Open-science Project and Status of the v2.0 Core Package}",
      journal = {\aj},
         year = 2018,
        month = sep,
       volume = {156},
       number = {3},
          eid = {123},
        pages = {123},
          doi = {10.3847/1538-3881/aabc4f},
archivePrefix = {arXiv},
       eprint = {1801.02634},
 primaryClass = {astro-ph.IM},
       adsurl = {https://ui.adsabs.harvard.edu/abs/2018AJ....156..123A}
}

@ARTICLE{astropy3,
       author = {{Astropy Collaboration} and {Price-Whelan}, Adrian M. and {Lim}, Pey Lian and {Earl}, Nicholas and {Starkman}, Nathaniel and {Bradley}, Larry and {Shupe}, David L. and {Patil}, Aarya A. and {Corrales}, Lia and {Brasseur}, C.~E. and {N{\"o}the}, Maximilian and {Donath}, Axel and {Tollerud}, Erik and {Morris}, Brett M. and {Ginsburg}, Adam and {Vaher}, Eero and {Weaver}, Benjamin A. and {Tocknell}, James and {Jamieson}, William and {van Kerkwijk}, Marten H. and {Robitaille}, Thomas P. and {Merry}, Bruce and {Bachetti}, Matteo and {G{\"u}nther}, H. Moritz and {Aldcroft}, Thomas L. and {Alvarado-Montes}, Jaime A. and {Archibald}, Anne M. and {B{\'o}di}, Attila and {Bapat}, Shreyas and {Barentsen}, Geert and {Baz{\'a}n}, Juanjo and {Biswas}, Manish and {Boquien}, M{\'e}d{\'e}ric and {Burke}, D.~J. and {Cara}, Daria and {Cara}, Mihai and {Conroy}, Kyle E. and {Conseil}, Simon and {Craig}, Matthew W. and {Cross}, Robert M. and {Cruz}, Kelle L. and {D'Eugenio}, Francesco and {Dencheva}, Nadia and {Devillepoix}, Hadrien A.~R. and {Dietrich}, J{\"o}rg P. and {Eigenbrot}, Arthur Davis and {Erben}, Thomas and {Ferreira}, Leonardo and {Foreman-Mackey}, Daniel and {Fox}, Ryan and {Freij}, Nabil and {Garg}, Suyog and {Geda}, Robel and {Glattly}, Lauren and {Gondhalekar}, Yash and {Gordon}, Karl D. and {Grant}, David and {Greenfield}, Perry and {Groener}, Austen M. and {Guest}, Steve and {Gurovich}, Sebastian and {Handberg}, Rasmus and {Hart}, Akeem and {Hatfield-Dodds}, Zac and {Homeier}, Derek and {Hosseinzadeh}, Griffin and {Jenness}, Tim and {Jones}, Craig K. and {Joseph}, Prajwel and {Kalmbach}, J. Bryce and {Karamehmetoglu}, Emir and {Ka{\l}uszy{\'n}ski}, Miko{\l}aj and {Kelley}, Michael S.~P. and {Kern}, Nicholas and {Kerzendorf}, Wolfgang E. and {Koch}, Eric W. and {Kulumani}, Shankar and {Lee}, Antony and {Ly}, Chun and {Ma}, Zhiyuan and {MacBride}, Conor and {Maljaars}, Jakob M. and {Muna}, Demitri and {Murphy}, N.~A. and {Norman}, Henrik and {O'Steen}, Richard and {Oman}, Kyle A. and {Pacifici}, Camilla and {Pascual}, Sergio and {Pascual-Granado}, J. and {Patil}, Rohit R. and {Perren}, Gabriel I. and {Pickering}, Timothy E. and {Rastogi}, Tanuj and {Roulston}, Benjamin R. and {Ryan}, Daniel F. and {Rykoff}, Eli S. and {Sabater}, Jose and {Sakurikar}, Parikshit and {Salgado}, Jes{\'u}s and {Sanghi}, Aniket and {Saunders}, Nicholas and {Savchenko}, Volodymyr and {Schwardt}, Ludwig and {Seifert-Eckert}, Michael and {Shih}, Albert Y. and {Jain}, Anany Shrey and {Shukla}, Gyanendra and {Sick}, Jonathan and {Simpson}, Chris and {Singanamalla}, Sudheesh and {Singer}, Leo P. and {Singhal}, Jaladh and {Sinha}, Manodeep and {Sip{\H{o}}cz}, Brigitta M. and {Spitler}, Lee R. and {Stansby}, David and {Streicher}, Ole and {{\v{S}}umak}, Jani and {Swinbank}, John D. and {Taranu}, Dan S. and {Tewary}, Nikita and {Tremblay}, Grant R. and {de Val-Borro}, Miguel and {Van Kooten}, Samuel J. and {Vasovi{\'c}}, Zlatan and {Verma}, Shresth and {de Miranda Cardoso}, Jos{\'e} Vin{\'\i}cius and {Williams}, Peter K.~G. and {Wilson}, Tom J. and {Winkel}, Benjamin and {Wood-Vasey}, W.~M. and {Xue}, Rui and {Yoachim}, Peter and {Zhang}, Chen and {Zonca}, Andrea and {Astropy Project Contributors}},
        title = "{The Astropy Project: Sustaining and Growing a Community-oriented Open-source Project and the Latest Major Release (v5.0) of the Core Package}",
      journal = {\apj},
         year = 2022,
        month = aug,
       volume = {935},
       number = {2},
          eid = {167},
        pages = {167},
          doi = {10.3847/1538-4357/ac7c74},
archivePrefix = {arXiv},
       eprint = {2206.14220},
 primaryClass = {astro-ph.IM},
       adsurl = {https://ui.adsabs.harvard.edu/abs/2022ApJ...935..167A}
}

@ARTICLE{Blondin06,
   author = {{Blondin}, S. and {Dessart}, L. and {Leibundgut}, B. and {Branch}, D. and {H{\"o}flich}, P. and {Tonry}, J.~L. and {Matheson}, T. and {Foley}, R.~J. and {Chornock}, R. and {Filippenko}, A.~V. and {Sollerman}, J. and {Spyromilio}, J. and {Kirshner}, R.~P. and {Wood-Vasey}, W.~M. and {Clocchiatti}, A. and {Aguilera}, C. and {Barris}, B. and {Becker}, A.~C. and {Challis}, P. and {Covarrubias}, R. and {Davis}, T.~M. and {Garnavich}, P. and {Hicken}, M. and {Jha}, S. and {Krisciunas}, K. and {Li}, W. and {Miceli}, A. and {Miknaitis}, G. and {Pignata}, G. and {Prieto}, J.~L. and {Rest}, A. and {Riess}, A.~G. and {Salvo}, M.~E. and {Schmidt}, B.~P. and {Smith}, R.~C. and {Stubbs}, C.~W. and {Suntzeff}, N.~B.},
    title = "{Using Line Profiles to Test the Fraternity of Type Ia Supernovae at High and Low Redshifts}",
  journal = {\aj},
   eprint = {arXiv:astro-ph/0510089},
     year = 2006,
    month = mar,
   volume = 131,
    pages = {1648-1666},
      doi = {10.1086/498724},
   adsurl = {http://adsabs.harvard.edu/abs/2006AJ....131.1648B}
}

@ARTICLE{Branch06,
   author = {{Branch}, D. and {Dang}, L.~C. and {Hall}, N. and {Ketchum}, W. and {Melakayil}, M. and {Parrent}, J. and {Troxel}, M.~A. and {Casebeer}, D. and {Jeffery}, D.~J. and {Baron}, E.},
    title = "{Comparative Direct Analysis of Type Ia Supernova Spectra. II. Maximum Light}",
  journal = {\pasp},
   eprint = {arXiv:astro-ph/0601048},
     year = 2006,
    month = apr,
   volume = 118,
    pages = {560-571},
      doi = {10.1086/502778},
   adsurl = {http://adsabs.harvard.edu/abs/2006PASP..118..560B}
}

@ARTICLE{Childress14,
   author = {{Childress}, M.~J. and {Wolf}, C. and {Zahid}, H.~J.},
    title = "{Ages of Type Ia supernovae over cosmic time}",
  journal = {\mnras},
archivePrefix = "arXiv",
   eprint = {1409.2951},
     year = 2014,
    month = dec,
   volume = 445,
    pages = {1898-1911},
      doi = {10.1093/mnras/stu1892},
   adsurl = {http://adsabs.harvard.edu/abs/2014MNRAS.445.1898C}
}

@ARTICLE{Coil00,
   author = {{Coil}, A.~L. and {Matheson}, T. and {Filippenko}, A.~V. and {Leonard}, D.~C. and {Tonry}, J. and {Riess}, A.~G. and {Challis}, P. and {Clocchiatti}, A. and {Garnavich}, P.~M. and {Hogan}, C.~J. and {Jha}, S. and {Kirshner}, R.~P. and {Leibundgut}, B. and {Phillips}, M.~M. and {Schmidt}, B.~P. and {Schommer}, R.~A. and {Smith}, R.~C. and {Soderberg}, A.~M. and {Spyromilio}, J. and {Stubbs}, C. and {Suntzeff}, N.~B. and {Woudt}, P.},
    title = "{Optical Spectra of Type IA Supernovae at Z=0.46 and Z=1.2}",
  journal = {\apjl},
   eprint = {arXiv:astro-ph/0009102},
     year = 2000,
    month = dec,
   volume = 544,
    pages = {L111-L114},
      doi = {10.1086/317311},
   adsurl = {http://adsabs.harvard.edu/abs/2000ApJ...544L.111C}
}

@ARTICLE{Conley11,
   author = {{Conley}, A. and {Guy}, J. and {Sullivan}, M. and {Regnault}, N. and {Astier}, P. and {Balland}, C. and {Basa}, S. and {Carlberg}, R.~G. and {Fouchez}, D. and {Hardin}, D. and {Hook}, I.~M. and {Howell}, D.~A. and {Pain}, R. and {Palanque-Delabrouille}, N. and {Perrett}, K.~M. and {Pritchet}, C.~J. and {Rich}, J. and {Ruhlmann-Kleider}, V. and {Balam}, D. and {Baumont}, S. and {Ellis}, R.~S. and {Fabbro}, S. and {Fakhouri}, H.~K. and {Fourmanoit}, N. and {Gonz{\'a}lez-Gait{\'a}n}, S. and {Graham}, M.~L. and {Hudson}, M.~J. and {Hsiao}, E. and {Kronborg}, T. and {Lidman}, C. and {Mourao}, A.~M. and {Neill}, J.~D. and {Perlmutter}, S. and {Ripoche}, P. and {Suzuki}, N. and {Walker}, E.~S.},
    title = "{Supernova Constraints and Systematic Uncertainties from the First Three Years of the Supernova Legacy Survey}",
  journal = {\apjs},
archivePrefix = "arXiv",
   eprint = {1104.1443},
 primaryClass = "astro-ph.CO",
     year = 2011,
    month = jan,
   volume = 192,
    pages = {1-+},
      doi = {10.1088/0067-0049/192/1/1},
   adsurl = {http://adsabs.harvard.edu/abs/2011ApJS..192....1C}
}

@ARTICLE{Foley11:vel,
   author = {{Foley}, R.~J. and {Kasen}, D.},
    title = "{Measuring Ejecta Velocity Improves Type Ia Supernova Distances}",
  journal = {\apj},
archivePrefix = "arXiv",
   eprint = {1011.4517},
 primaryClass = "astro-ph.CO",
     year = 2011,
    month = mar,
   volume = 729,
    pages = {55-+},
      doi = {10.1088/0004-637X/729/1/55},
   adsurl = {http://adsabs.harvard.edu/abs/2011ApJ...729...55F}
}

@ARTICLE{Foley12:vel,
   author = {{Foley}, R.~J.},
    title = "{The Relation between Ejecta Velocity, Intrinsic Color, and Host-galaxy Mass for High-redshift Type Ia Supernovae}",
  journal = {\apj},
archivePrefix = "arXiv",
   eprint = {1202.0003},
 primaryClass = "astro-ph.CO",
     year = 2012,
    month = apr,
   volume = 748,
      eid = {127},
    pages = {127},
      doi = {10.1088/0004-637X/748/2/127},
   adsurl = {http://adsabs.harvard.edu/abs/2012ApJ...748..127F}
}

@ARTICLE{Foley12:sdss,
   author = {{Foley}, R.~J. and {Filippenko}, A.~V. and {Kessler}, R. and {Bassett}, B. and {Frieman}, J.~A. and {Garnavich}, P.~M. and {Jha}, S.~W. and {Konishi}, K. and {Lampeitl}, H. and {Riess}, A.~G. and {Sako}, M. and {Schneider}, D.~P. and {Sollerman}, J. and {Smith}, M.},
    title = "{A Mismatch in the Ultraviolet Spectra between Low-redshift and Intermediate-redshift Type Ia Supernovae as a Possible Systematic Uncertainty for Supernova Cosmology}",
  journal = {\aj},
archivePrefix = "arXiv",
   eprint = {1010.2749},
 primaryClass = "astro-ph.CO",
     year = 2012,
    month = may,
   volume = 143,
      eid = {113},
    pages = {113},
      doi = {10.1088/0004-6256/143/5/113},
   adsurl = {http://adsabs.harvard.edu/abs/2012AJ....143..113F}
}

@ARTICLE{Hachinger13,
   author = {{Hachinger}, S. and {Mazzali}, P.~A. and {Sullivan}, M. and {Ellis}, R.~S. and {Maguire}, K. and {Gal-Yam}, A. and {Howell}, D.~A. and {Nugent}, P.~E. and {Baron}, E. and {Cooke}, J. and {Arcavi}, I. and {Bersier}, D. and {Dilday}, B. and {James}, P.~A. and {Kasliwal}, M.~M. and {Kulkarni}, S.~R. and {Ofek}, E.~O. and {Laher}, R.~R. and {Parrent}, J. and {Surace}, J. and {Yaron}, O. and {Walker}, E.~S.},
    title = "{The UV/optical spectra of the Type Ia supernova SN 2010jn: a bright supernova with outer layers rich in iron-group elements}",
  journal = {\mnras},
archivePrefix = "arXiv",
   eprint = {1208.1267},
 primaryClass = "astro-ph.SR",
     year = 2013,
    month = mar,
   volume = 429,
    pages = {2228-2248},
      doi = {10.1093/mnras/sts492},
   adsurl = {http://adsabs.harvard.edu/abs/2013MNRAS.429.2228H}
}

@ARTICLE{Hoflich98,
   author = {{H\"{o}flich}, P. and {Wheeler}, J.~C. and {Thielemann}, F.-K.},
    title = "{Type IA Supernovae: Influence of the Initial Composition on the Nucleosynthesis, Light Curves, and Spectra and Consequences for the Determination of Omega M and Lambda}",
  journal = {\apj},
   eprint = {arXiv:astro-ph/9709233},
     year = 1998,
    month = mar,
   volume = 495,
    pages = {617-+},
      doi = {10.1086/305327},
   adsurl = {http://adsabs.harvard.edu/abs/1998ApJ...495..617H}
}

@ARTICLE{Jones18:ps1,
   author = {{Jones}, D.~O. and {Scolnic}, D.~M. and {Riess}, A.~G. and {Rest}, A. and {Kirshner}, R.~P. and {Berger}, E. and {Kessler}, R. and {Pan}, Y.-C. and {Foley}, R.~J. and {Chornock}, R. and {Ortega}, C.~A. and {Challis}, P.~J. and {Burgett}, W.~S. and {Chambers}, K.~C. and {Draper}, P.~W. and {Flewelling}, H. and {Huber}, M.~E. and {Kaiser}, N. and {Kudritzki}, R.-P. and {Metcalfe}, N. and {Tonry}, J. and {Wainscoat}, R.~J. and {Waters}, C. and {Gall}, E.~E.~E. and {Kotak}, R. and {McCrum}, M. and {Smartt}, S.~J. and {Smith}, K.~W.},
    title = "{Measuring Dark Energy Properties with Photometrically Classified Pan-STARRS Supernovae. II. Cosmological Parameters}",
  journal = {\apj},
archivePrefix = "arXiv",
   eprint = {1710.00846},
     year = 2018,
    month = apr,
   volume = 857,
      eid = {51},
    pages = {51},
      doi = {10.3847/1538-4357/aab6b1},
   adsurl = {http://adsabs.harvard.edu/abs/2018ApJ...857...51J}
}

@ARTICLE{Kelly10,
   author = {{Kelly}, P.~L. and {Hicken}, M. and {Burke}, D.~L. and {Mandel}, K.~S. and {Kirshner}, R.~P.},
    title = "{Hubble Residuals of Nearby Type Ia Supernovae are Correlated with Host Galaxy Masses}",
  journal = {\apj},
archivePrefix = "arXiv",
   eprint = {0912.0929},
 primaryClass = "astro-ph.CO",
     year = 2010,
    month = jun,
   volume = 715,
    pages = {743-756},
      doi = {10.1088/0004-637X/715/2/743},
   adsurl = {http://adsabs.harvard.edu/abs/2010ApJ...715..743K}
}

@ARTICLE{Lampeitl10:host,
   author = {{Lampeitl}, H. and {Smith}, M. and {Nichol}, R.~C. and {Bassett}, B. and {Cinabro}, D. and {Dilday}, B. and {Foley}, R.~J. and {Frieman}, J.~A. and {Garnavich}, P.~M. and {Goobar}, A. and {Im}, M. and {Jha}, S.~W. and {Marriner}, J. and {Miquel}, R. and {Nordin}, J. and {{\"O}stman}, L. and {Riess}, A.~G. and {Sako}, M. and {Schneider}, D.~P. and {Sollerman}, J. and {Stritzinger}, M.},
    title = "{The Effect of Host Galaxies on Type Ia Supernovae in the SDSS-II Supernova Survey}",
  journal = {\apj},
archivePrefix = "arXiv",
   eprint = {1005.4687},
 primaryClass = "astro-ph.CO",
     year = 2010,
    month = oct,
   volume = 722,
    pages = {566-576},
      doi = {10.1088/0004-637X/722/1/566},
   adsurl = {http://adsabs.harvard.edu/abs/2010ApJ...722..566L}
}

@ARTICLE{Lentz00,
   author = {{Lentz}, E.~J. and {Baron}, E. and {Branch}, D. and {Hauschildt}, P.~H. and {Nugent}, P.~E.},
    title = "{Metallicity Effects in Non-LTE Model Atmospheres of Type IA Supernovae}",
  journal = {\apj},
   eprint = {arXiv:astro-ph/9906016},
     year = 2000,
    month = feb,
   volume = 530,
    pages = {966-976},
      doi = {10.1086/308400},
   adsurl = {http://adsabs.harvard.edu/abs/2000ApJ...530..966L}
}

@ARTICLE{Maoz14,
   author = {{Maoz}, D. and {Mannucci}, F. and {Nelemans}, G.},
    title = "{Observational Clues to the Progenitors of Type Ia Supernovae}",
  journal = {\araa},
archivePrefix = "arXiv",
   eprint = {1312.0628},
     year = 2014,
    month = aug,
   volume = 52,
    pages = {107-170},
      doi = {10.1146/annurev-astro-082812-141031},
   adsurl = {http://adsabs.harvard.edu/abs/2014ARA%26A..52..107M}
}

@ARTICLE{Mazzali14,
   author = {{Mazzali}, P.~A. and {Sullivan}, M. and {Hachinger}, S. and {Ellis}, R.~S. and {Nugent}, P.~E. and {Howell}, D.~A. and {Gal-Yam}, A. and {Maguire}, K. and {Cooke}, J. and {Thomas}, R. and {Nomoto}, K. and {Walker}, E.~S.},
    title = "{Hubble Space Telescope spectra of the Type Ia supernova SN 2011fe: a tail of low-density, high-velocity material with Z $<$ Z$_{\sun}$}",
  journal = {\mnras},
archivePrefix = "arXiv",
   eprint = {1305.2356},
 primaryClass = "astro-ph.CO",
     year = 2014,
    month = feb,
   volume = 439,
    pages = {1959-1979},
      doi = {10.1093/mnras/stu077},
   adsurl = {http://adsabs.harvard.edu/abs/2014MNRAS.439.1959M}
}

@ARTICLE{Nugent95,
   author = {{Nugent}, P. and {Phillips}, M. and {Baron}, E. and {Branch}, D. and {Hauschildt}, P.},
    title = "{Evidence for a Spectroscopic Sequence among Type 1a Supernovae}",
  journal = {\apjl},
   eprint = {arXiv:astro-ph/9510004},
     year = 1995,
    month = dec,
   volume = 455,
    pages = {L147},
      doi = {10.1086/309846},
   adsurl = {http://adsabs.harvard.edu/abs/1995ApJ...455L.147N}
}

@ARTICLE{Perlmutter99,
   author = {{Perlmutter}, S. and {Aldering}, G. and {Goldhaber}, G. and {Knop}, R.~A. and {Nugent}, P. and {Castro}, P.~G. and {Deustua}, S. and {Fabbro}, S. and {Goobar}, A. and {Groom}, D.~E. and {Hook}, I.~M. and {Kim}, A.~G. and {Kim}, M.~Y. and {Lee}, J.~C. and {Nunes}, N.~J. and {Pain}, R. and {Pennypacker}, C.~R. and {Quimby}, R. and {Lidman}, C. and {Ellis}, R.~S. and {Irwin}, M. and {McMahon}, R.~G. and {Ruiz-Lapuente}, P. and {Walton}, N. and {Schaefer}, B. and {Boyle}, B.~J. and {Filippenko}, A.~V. and {Matheson}, T. and {Fruchter}, A.~S. and {Panagia}, N. and {Newberg}, H.~J.~M. and {Couch}, W.~J.},
    title = "{Measurements of Omega and Lambda from 42 High-Redshift Supernovae}",
  journal = {\apj},
   eprint = {arXiv:astro-ph/9812133},
     year = 1999,
    month = jun,
   volume = 517,
    pages = {565-586},
      doi = {10.1086/307221},
   adsurl = {http://adsabs.harvard.edu/abs/1999ApJ...517..565P}
}

@ARTICLE{Phillips93,
   author = {{Phillips}, M.~M.},
    title = "{The absolute magnitudes of Type IA supernovae}",
  journal = {\apjl},
     year = 1993,
    month = aug,
   volume = 413,
    pages = {L105-L108},
      doi = {10.1086/186970},
   adsurl = {http://adsabs.harvard.edu/abs/1993ApJ...413L.105P}
}

@ARTICLE{Phillips99,
   author = {{Phillips}, M.~M. and {Lira}, P. and {Suntzeff}, N.~B. and {Schommer}, R.~A. and {Hamuy}, M. and {Maza}, J.},
    title = "{The Reddening-Free Decline Rate Versus Luminosity Relationship for Type IA Supernovae}",
  journal = {\aj},
   eprint = {arXiv:astro-ph/9907052},
     year = 1999,
    month = oct,
   volume = 118,
    pages = {1766-1776},
      doi = {10.1086/301032},
   adsurl = {http://adsabs.harvard.edu/abs/1999AJ....118.1766P}
}

@ARTICLE{Riess98:lambda,
   author = {{Riess}, A.~G. and {Filippenko}, A.~V. and {Challis}, P. and {Clocchiatti}, A. and {Diercks}, A. and {Garnavich}, P.~M. and {Gilliland}, R.~L. and {Hogan}, C.~J. and {Jha}, S. and {Kirshner}, R.~P. and {Leibundgut}, B. and {Phillips}, M.~M. and {Reiss}, D. and {Schmidt}, B.~P. and {Schommer}, R.~A. and {Smith}, R.~C. and {Spyromilio}, J. and {Stubbs}, C. and {Suntzeff}, N.~B. and {Tonry}, J.},
    title = "{Observational Evidence from Supernovae for an Accelerating Universe and a Cosmological Constant}",
  journal = {\aj},
     year = 1998,
    month = sep,
   volume = 116,
    pages = {1009-1038},
   adsurl = {http://adsabs.harvard.edu/cgi-bin/nph-bib_query?bibcode=1998AJ....116.1009R&db_key=AST}
}

@ARTICLE{Riess06,
   author = {{Riess}, A.~G. and {Livio}, M.},
    title = "{The First Type Ia Supernovae: An Empirical Approach to Taming Evolutionary Effects in Dark Energy Surveys from SNe Ia at z{\gt}2}",
  journal = {\apj},
   eprint = {astro-ph/0601319},
     year = 2006,
    month = sep,
   volume = 648,
    pages = {884-889},
      doi = {10.1086/504791},
   adsurl = {http://adsabs.harvard.edu/abs/2006ApJ...648..884R}
}

@ARTICLE{Sauer08,
   author = {{Sauer}, D.~N. and {Mazzali}, P.~A. and {Blondin}, S. and {Stehle}, M. and {Benetti}, S. and {Challis}, P. and {Filippenko}, A.~V. and {Kirshner}, R.~P. and {Li}, W. and {Matheson}, T.},
    title = "{Properties of the ultraviolet flux of Type Ia supernovae: an analysis with synthetic spectra of SN 2001ep and SN 2001eh}",
  journal = {\mnras},
archivePrefix = "arXiv",
   eprint = {0803.0871},
     year = 2008,
    month = dec,
   volume = 391,
    pages = {1605-1618},
      doi = {10.1111/j.1365-2966.2008.14018.x},
   adsurl = {http://adsabs.harvard.edu/abs/2008MNRAS.391.1605S}
}

@ARTICLE{Scolnic18:ps1,
   author = {{Scolnic}, D.~M. and {Jones}, D.~O. and {Rest}, A. and {Pan}, Y.~C. and {Chornock}, R. and {Foley}, R.~J. and {Huber}, M.~E. and {Kessler}, R. and {Narayan}, G. and {Riess}, A.~G. and {Rodney}, S. and {Berger}, E. and {Brout}, D.~J. and {Challis}, P.~J. and {Drout}, M. and {Finkbeiner}, D. and {Lunnan}, R. and {Kirshner}, R.~P. and {Sanders}, N.~E. and {Schlafly}, E. and {Smartt}, S. and {Stubbs}, C.~W. and {Tonry}, J. and {Wood-Vasey}, W.~M. and {Foley}, M. and {Hand}, J. and {Johnson}, E. and {Burgett}, W.~S. and {Chambers}, K.~C. and {Draper}, P.~W. and {Hodapp}, K.~W. and {Kaiser}, N. and {Kudritzki}, R.~P. and {Magnier}, E.~A. and {Metcalfe}, N. and {Bresolin}, F. and {Gall}, E. and {Kotak}, R. and {McCrum}, M. and {Smith}, K.~W.},
    title = "{The Complete Light-curve Sample of Spectroscopically Confirmed SNe Ia from Pan-STARRS1 and Cosmological Constraints from the Combined Pantheon Sample}",
  journal = {\apj},
archivePrefix = "arXiv",
   eprint = {1710.00845},
     year = 2018,
    month = jun,
   volume = 859,
      eid = {101},
    pages = {101},
      doi = {10.3847/1538-4357/aab9bb},
   adsurl = {http://adsabs.harvard.edu/abs/2018ApJ...859..101S}
}

@ARTICLE{Siebert19,
       author = {{Siebert}, M.~R. and {Foley}, R.~J. and {Jones}, D.~O. and {Angulo}, R. and
         {Davis}, K. and {Duarte}, A. and {Strasburger}, E. and {Conlon}, M. and
         {Kazmi}, N. and {Nishimoto}, R. and {Schubert}, M. and {Sun}, L. and
         {Tippens}, R.},
        title = "{Investigating the diversity of Type Ia supernova spectra with the open-source relational data base KAEPORA}",
      journal = {\mnras},
         year = "2019",
        month = "Jul",
       volume = {486},
       number = {4},
        pages = {5785-5808},
          doi = {10.1093/mnras/stz1209},
archivePrefix = {arXiv},
       eprint = {1905.02204},
 primaryClass = {astro-ph.HE},
       adsurl = {https://ui.adsabs.harvard.edu/abs/2019MNRAS.486.5785S}
}

@ARTICLE{Siebert20a,
       author = {{Siebert}, M.~R. and {Foley}, R.~J. and {Jones}, D.~O. and
         {Davis}, K.~W.},
        title = "{A Possible Distance Bias for Type Ia Supernovae with Different Ejecta Velocities}",
      journal = {arXiv e-prints},
         year = 2020,
        month = feb,
          eid = {arXiv:2002.09490},
        pages = {arXiv:2002.09490},
archivePrefix = {arXiv},
       eprint = {2002.09490},
 primaryClass = {astro-ph.GA},
       adsurl = {https://ui.adsabs.harvard.edu/abs/2020arXiv200209490S}
}

@ARTICLE{Sternberg11,
   author = {{Sternberg}, A. and {Gal-Yam}, A. and {Simon}, J.~D. and {Leonard}, D.~C. and {Quimby}, R.~M. and {Phillips}, M.~M. and {Morrell}, N. and {Thompson}, I.~B. and {Ivans}, I. and {Marshall}, J.~L. and {Filippenko}, A.~V. and {Marcy}, G.~W. and {Bloom}, J.~S. and {Patat}, F. and {Foley}, R.~J. and {Yong}, D. and {Penprase}, B.~E. and {Beeler}, D.~J. and {Allende Prieto}, C. and {Stringfellow}, G.~S.},
    title = "{Circumstellar Material in Type Ia Supernovae via Sodium Absorption Features}",
  journal = {Science},
archivePrefix = "arXiv",
   eprint = {1108.3664},
 primaryClass = "astro-ph.HE",
     year = 2011,
    month = aug,
   volume = 333,
    pages = {856-},
      doi = {10.1126/science.1203836},
   adsurl = {http://adsabs.harvard.edu/abs/2011Sci...333..856S}
}

@ARTICLE{Sullivan10,
   author = {{Sullivan}, M. and {Conley}, A. and {Howell}, D.~A. and {Neill}, J.~D. and {Astier}, P. and {Balland}, C. and {Basa}, S. and {Carlberg}, R.~G. and {Fouchez}, D. and {Guy}, J. and {Hardin}, D. and {Hook}, I.~M. and {Pain}, R. and {Palanque-Delabrouille}, N. and {Perrett}, K.~M. and {Pritchet}, C.~J. and {Regnault}, N. and {Rich}, J. and {Ruhlmann-Kleider}, V. and {Baumont}, S. and {Hsiao}, E. and {Kronborg}, T. and {Lidman}, C. and {Perlmutter}, S. and {Walker}, E.~S.},
    title = "{The dependence of Type Ia Supernovae luminosities on their host galaxies}",
  journal = {\mnras},
archivePrefix = "arXiv",
   eprint = {1003.5119},
 primaryClass = "astro-ph.CO",
     year = 2010,
    month = aug,
   volume = 406,
    pages = {782-802},
      doi = {10.1111/j.1365-2966.2010.16731.x},
   adsurl = {http://adsabs.harvard.edu/abs/2010MNRAS.406..782S}
}

@INCOLLECTION{Taubenberger17,
       author = {{Taubenberger}, Stefan},
        title = "{The Extremes of Thermonuclear Supernovae}",
    booktitle = {Handbook of Supernovae},
         year = 2017,
       editor = {{Alsabti}, Athem W. and {Murdin}, Paul},
        pages = {317},
    publisher = {Springer Cham},
          doi = {10.1007/978-3-319-21846-5_37},
       adsurl = {https://ui.adsabs.harvard.edu/abs/2017hsn..book..317T}
}

@ARTICLE{Tripp98,
   author = {{Tripp}, R.},
    title = "{A two-parameter luminosity correction for Type IA supernovae}",
  journal = {\aap},
     year = 1998,
    month = mar,
   volume = 331,
    pages = {815-820},
   adsurl = {http://adsabs.harvard.edu/abs/1998A%26A...331..815T}
}

@ARTICLE{Scolnic22,
       author = {{Scolnic}, Dan and {Brout}, Dillon and {Carr}, Anthony and {Riess}, Adam G. and {Davis}, Tamara M. and {Dwomoh}, Arianna and {Jones}, David O. and {Ali}, Noor and {Charvu}, Pranav and {Chen}, Rebecca and {Peterson}, Erik R. and {Popovic}, Brodie and {Rose}, Benjamin M. and {Wood}, Charlotte M. and {Brown}, Peter J. and {Chambers}, Ken and {Coulter}, David A. and {Dettman}, Kyle G. and {Dimitriadis}, Georgios and {Filippenko}, Alexei V. and {Foley}, Ryan J. and {Jha}, Saurabh W. and {Kilpatrick}, Charles D. and {Kirshner}, Robert P. and {Pan}, Yen-Chen and {Rest}, Armin and {Rojas-Bravo}, Cesar and {Siebert}, Matthew R. and {Stahl}, Benjamin E. and {Zheng}, WeiKang},
        title = "{The Pantheon+ Analysis: The Full Data Set and Light-curve Release}",
      journal = {\apj},
         year = 2022,
        month = oct,
       volume = {938},
       number = {2},
          eid = {113},
        pages = {113},
          doi = {10.3847/1538-4357/ac8b7a},
archivePrefix = {arXiv},
       eprint = {2112.03863},
 primaryClass = {astro-ph.CO},
       adsurl = {https://ui.adsabs.harvard.edu/abs/2022ApJ...938..113S}
}

@ARTICLE{Jakobsen2022,
       author = {{Jakobsen}, P. and {Ferruit}, P. and {Alves de Oliveira}, C. and {Arribas}, S. and {Bagnasco}, G. and {Barho}, R. and {Beck}, T.~L. and {Birkmann}, S. and {B{\"o}ker}, T. and {Bunker}, A.~J. and {Charlot}, S. and {de Jong}, P. and {de Marchi}, G. and {Ehrenwinkler}, R. and {Falcolini}, M. and {Fels}, R. and {Franx}, M. and {Franz}, D. and {Funke}, M. and {Giardino}, G. and {Gnata}, X. and {Holota}, W. and {Honnen}, K. and {Jensen}, P.~L. and {Jentsch}, M. and {Johnson}, T. and {Jollet}, D. and {Karl}, H. and {Kling}, G. and {K{\"o}hler}, J. and {Kolm}, M. -G. and {Kumari}, N. and {Lander}, M.~E. and {Lemke}, R. and {L{\'o}pez-Caniego}, M. and {L{\"u}tzgendorf}, N. and {Maiolino}, R. and {Manjavacas}, E. and {Marston}, A. and {Maschmann}, M. and {Maurer}, R. and {Messerschmidt}, B. and {Moseley}, S.~H. and {Mosner}, P. and {Mott}, D.~B. and {Muzerolle}, J. and {Pirzkal}, N. and {Pittet}, J. -F. and {Plitzke}, A. and {Posselt}, W. and {Rapp}, B. and {Rauscher}, B.~J. and {Rawle}, T. and {Rix}, H. -W. and {R{\"o}del}, A. and {Rumler}, P. and {Sabbi}, E. and {Salvignol}, J. -C. and {Schmid}, T. and {Sirianni}, M. and {Smith}, C. and {Strada}, P. and {te Plate}, M. and {Valenti}, J. and {Wettemann}, T. and {Wiehe}, T. and {Wiesmayer}, M. and {Willott}, C.~J. and {Wright}, R. and {Zeidler}, P. and {Zincke}, C.},
        title = "{The Near-Infrared Spectrograph (NIRSpec) on the James Webb Space Telescope. I. Overview of the instrument and its capabilities}",
      journal = {\aap},
         year = 2022,
        month = may,
       volume = {661},
          eid = {A80},
        pages = {A80},
          doi = {10.1051/0004-6361/202142663},
archivePrefix = {arXiv},
       eprint = {2202.03305},
 primaryClass = {astro-ph.IM},
       adsurl = {https://ui.adsabs.harvard.edu/abs/2022A&A...661A..80J}
}

@ARTICLE{Rigby2022,
       author = {{Rigby}, Jane and {Perrin}, Marshall and {McElwain}, Michael and {Kimble}, Randy and {Friedman}, Scott and {Lallo}, Matt and {Doyon}, Ren{\'e} and {Feinberg}, Lee and {Ferruit}, Pierre and {Glasse}, Alistair and et al.},
        title = "{Characterization of JWST science performance from commissioning}",
      journal = {arXiv e-prints},
         year = 2022,
        month = jul,
          eid = {arXiv:2207.05632},
        pages = {arXiv:2207.05632},
archivePrefix = {arXiv},
       eprint = {2207.05632},
 primaryClass = {astro-ph.IM},
       adsurl = {https://ui.adsabs.harvard.edu/abs/2022arXiv220705632R}
}

@INPROCEEDINGS{Birkmann2022,
       author = {{Birkmann}, Stephan M. and {Giardino}, Giovanna and {Sirianni}, Marco and {Ferruit}, Pierre and {Rauscher}, Bernhard and {Alves de Oliveira}, Catarina and {B{\"o}ker}, Torsten and {Kumari}, Nimisha and {L{\"u}tzgendorf}, Nora and {Manjavacas}, Elena and {Proffitt}, Charles and {Rawle}, Timothy D. and {te Plate}, Maurice and {Zeidler}, Peter},
        title = "{The in-flight noise performance of the JWST/NIRSpec detector system}",
    booktitle = {Space Telescopes and Instrumentation 2022: Optical, Infrared, and Millimeter Wave},
         year = 2022,
       editor = {{Coyle}, Laura E. and {Matsuura}, Shuji and {Perrin}, Marshall D.},
       series = {Society of Photo-Optical Instrumentation Engineers (SPIE) Conference Series},
       volume = {12180},
        month = aug,
          eid = {121802P},
        pages = {121802P},
          doi = {10.1117/12.2629545},
archivePrefix = {arXiv},
       eprint = {2208.12686},
 primaryClass = {astro-ph.IM},
       adsurl = {https://ui.adsabs.harvard.edu/abs/2022SPIE12180E..2PB}
}

@software{Bushouse_JWST_Calibration_Pipeline_2024,
author = {Bushouse, Howard and Eisenhamer, Jonathan and Dencheva, Nadia and Davies, James and Greenfield, Perry and Morrison, Jane and Hodge, Phil and Simon, Bernie and Grumm, David and Droettboom, Michael and Slavich, Edward and Sosey, Megan and Pauly, Tyler and Miller, Todd and Jedrzejewski, Robert and Hack, Warren and Davis, David and Crawford, Steven and Law, David and Gordon, Karl and Regan, Michael and Cara, Mihai and MacDonald, Ken and Bradley, Larry and Shanahan, Clare and Jamieson, William and Teodoro, Mairan and Williams, Thomas},
doi = {10.5281/zenodo.7038885},
month = {9},
title = {{JWST Calibration Pipeline}},
url = {https://github.com/spacetelescope/jwst},
version = {1.14.0},
year = {2024}
}

@ARTICLE{Goldwasser22,
       author = {{Goldwasser}, S. and {Yaron}, O. and {Sass}, A. and {Irani}, I. and {Gal-Yam}, A. and {Howell}, D.~A.},
        title = "{The Next Generation SuperFit (NGSF) tool is now available for online execution on WISeREP}",
      journal = {Transient Name Server AstroNote},
         year = 2022,
        month = sep,
       volume = {191},
        pages = {1},
       adsurl = {https://ui.adsabs.harvard.edu/abs/2022TNSAN.191....1G}
}

@ARTICLE{Eisenstein23,
       author = {{Eisenstein}, Daniel J. and {Willott}, Chris and {Alberts}, Stacey and {Arribas}, Santiago and {Bonaventura}, Nina and {Bunker}, Andrew J. and {Cameron}, Alex J. and {Carniani}, Stefano and {Charlot}, Stephane and {Curtis-Lake}, Emma and {D'Eugenio}, Francesco and {Endsley}, Ryan and {Ferruit}, Pierre and {Giardino}, Giovanna and {Hainline}, Kevin and {Hausen}, Ryan and {Jakobsen}, Peter and {Johnson}, Benjamin D. and {Maiolino}, Roberto and {Rieke}, Marcia and {Rieke}, George and {Rix}, Hans-Walter and {Robertson}, Brant and {Stark}, Daniel P. and {Tacchella}, Sandro and {Williams}, Christina C. and {Willmer}, Christopher N.~A. and {Baker}, William M. and {Baum}, Stefi and {Bhatawdekar}, Rachana and {Boyett}, Kristan and {Chen}, Zuyi and {Chevallard}, Jacopo and {Circosta}, Chiara and {Curti}, Mirko and {Danhaive}, A. Lola and {DeCoursey}, Christa and {de Graaff}, Anna and {Dressler}, Alan and {Egami}, Eiichi and {Helton}, Jakob M. and {Hviding}, Raphael E. and {Ji}, Zhiyuan and {Jones}, Gareth C. and {Kumari}, Nimisha and {L{\"u}tzgendorf}, Nora and {Laseter}, Isaac and {Looser}, Tobias J. and {Lyu}, Jianwei and {Maseda}, Michael V. and {Nelson}, Erica and {Parlanti}, Eleonora and {Perna}, Michele and {Pusk{\'a}s}, D{\'a}vid and {Rawle}, Tim and {Rodr{\'\i}guez Del Pino}, Bruno and {Sandles}, Lester and {Saxena}, Aayush and {Scholtz}, Jan and {Sharpe}, Katherine and {Shivaei}, Irene and {Silcock}, Maddie S. and {Simmonds}, Charlotte and {Skarbinski}, Maya and {Smit}, Renske and {Stone}, Meredith and {Suess}, Katherine A. and {Sun}, Fengwu and {Tang}, Mengtao and {Topping}, Michael W. and {{\"U}bler}, Hannah and {Villanueva}, Natalia C. and {Wallace}, Imaan E.~B. and {Whitler}, Lily and {Witstok}, Joris and {Woodrum}, Charity},
        title = "{Overview of the JWST Advanced Deep Extragalactic Survey (JADES)}",
      journal = {arXiv e-prints},
         year = 2023,
        month = jun,
          eid = {arXiv:2306.02465},
        pages = {arXiv:2306.02465},
          doi = {10.48550/arXiv.2306.02465},
archivePrefix = {arXiv},
       eprint = {2306.02465},
 primaryClass = {astro-ph.GA},
       adsurl = {https://ui.adsabs.harvard.edu/abs/2023arXiv230602465E}
}

@ARTICLE{Decoursey24,
       author = {{DeCoursey}, Christa and {Egami}, Eiichi and {Pierel}, Justin D.~R. and {Sun}, Fengwu and {Rest}, Armin and {Coulter}, David A. and {Engesser}, Michael and {Siebert}, Matthew R. and {Hainline}, Kevin N. and {Johnson}, Benjamin D. and {Bunker}, Andrew J. and {Cargile}, Phillip A. and {Charlot}, Stephane and {Chen}, Wenlei and {Curti}, Mirko and {DeFour-Remy}, Shea and {Eisenstein}, Daniel J. and {Fox}, Ori D. and {Gezari}, Suvi and {Gomez}, Sebastian and {Jencson}, Jacob and {Joshi}, Bhavin A. and {Khairnar}, Sanvi and {Lyu}, Jianwei and {Maiolino}, Roberto and {Moriya}, Takashi J. and {Quimby}, Robert M. and {Rieke}, George H. and {Rieke}, Marcia J. and {Robertson}, Brant and {Shahbandeh}, Melissa and {Strolger}, Louis-Gregory and {Tacchella}, Sandro and {Wang}, Qinan and {Williams}, Christina C. and {Willmer}, Christopher N.~A. and {Willott}, Chris and {Zenati}, Yossef},
        title = "{The JADES Transient Survey: Discovery and Classification of Supernovae in the JADES Deep Field}",
      journal = {arXiv e-prints},
         year = 2024,
        month = jun,
          eid = {arXiv:2406.05060},
        pages = {arXiv:2406.05060},
archivePrefix = {arXiv},
       eprint = {2406.05060},
 primaryClass = {astro-ph.HE},
       adsurl = {https://ui.adsabs.harvard.edu/abs/2024arXiv240605060D}
}

@ARTICLE{Pierel24,
       author = {{Pierel}, J.~D.~R. and {Engesser}, M. and {Coulter}, D.~A. and {Decoursey}, C. and {Siebert}, M. and {Rest}, A. and {Egami}, E. and {Chen}, W. and {Fox}, O.~D. and {Jones}, D.~O. and {Joshi}, B.~A. and {Moriya}, T.~J. and {Zenati}, Y. and {Bunker}, A.~J. and {Cargile}, P.~A. and {Curti}, M. and {Eisenstein}, D.~J. and {Gezari}, S. and {Gomez}, S. and {Guolo}, M. and {Johnson}, B.~D. and {Karmen}, M. and {Maiolino}, R. and {Quimby}, Robert M. and {Robertson}, B. and {Shahbandeh}, M. and {Strolger}, L.~G. and {Sun}, F. and {Wang}, Q. and {Wevers}, T.},
        title = "{Discovery of An Apparent Red, High-Velocity Type Ia Supernova at z = 2.9 with JWST}",
      journal = {arXiv e-prints},
         year = 2024,
        month = jun,
          eid = {arXiv:2406.05089},
        pages = {arXiv:2406.05089},
archivePrefix = {arXiv},
       eprint = {2406.05089},
 primaryClass = {astro-ph.GA},
       adsurl = {https://ui.adsabs.harvard.edu/abs/2024arXiv240605089P}
}

@ARTICLE{Adame25,
       author = {{Adame}, A.~G. and {Aguilar}, J. and {Ahlen}, S. and {Alam}, S. and {Alexander}, D.~M. and {Alvarez}, M. and {Alves}, O. and {Anand}, A. and {Andrade}, U. and {Armengaud}, E. and {Avila}, S. and {Aviles}, A. and {Awan}, H. and {Bahr-Kalus}, B. and {Bailey}, S. and {Baltay}, C. and {Bault}, A. and {Behera}, J. and {BenZvi}, S. and {Bera}, A. and {Beutler}, F. and {Bianchi}, D. and {Blake}, C. and {Blum}, R. and {Brieden}, S. and {Brodzeller}, A. and {Brooks}, D. and {Buckley-Geer}, E. and {Burtin}, E. and {Calderon}, R. and {Canning}, R. and {Carnero Rosell}, A. and {Cereskaite}, R. and {Cervantes-Cota}, J.~L. and {Chabanier}, S. and {Chaussidon}, E. and {Chaves-Montero}, J. and {Chen}, S. and {Chen}, X. and {Claybaugh}, T. and {Cole}, S. and {Cuceu}, A. and {Davis}, T.~M. and {Dawson}, K. and {de la Macorra}, A. and {de Mattia}, A. and {Deiosso}, N. and {Dey}, A. and {Dey}, B. and {Ding}, Z. and {Doel}, P. and {Edelstein}, J. and {Eftekharzadeh}, S. and {Eisenstein}, D.~J. and {Elliott}, A. and {Fagrelius}, P. and {Fanning}, K. and {Ferraro}, S. and {Ereza}, J. and {Findlay}, N. and {Flaugher}, B. and {Font-Ribera}, A. and {Forero-S{\'a}nchez}, D. and {Forero-Romero}, J.~E. and {Frenk}, C.~S. and {Garcia-Quintero}, C. and {Gazta{\~n}aga}, E. and {Gil-Mar{\'\i}n}, H. and {Gontcho a Gontcho}, S. and {Gonzalez-Morales}, A.~X. and {Gonzalez-Perez}, V. and {Gordon}, C. and {Green}, D. and {Gruen}, D. and {Gsponer}, R. and {Gutierrez}, G. and {Guy}, J. and {Hadzhiyska}, B. and {Hahn}, C. and {Hanif}, M.~M.~S. and {Herrera-Alcantar}, H.~K. and {Honscheid}, K. and {Howlett}, C. and {Huterer}, D. and {Ir{\v{s}}i{\v{c}}}, V. and {Ishak}, M. and {Juneau}, S. and {Kara{\c{c}}ayl{\i}}, N.~G. and {Kehoe}, R. and {Kent}, S. and {Kirkby}, D. and {Kremin}, A. and {Krolewski}, A. and {Lai}, Y. and {Lan}, T.-W. and {Landriau}, M. and {Lang}, D. and {Lasker}, J. and {Le Goff}, J.~M. and {Le Guillou}, L. and {Leauthaud}, A. and {Levi}, M.~E. and {Li}, T.~S. and {Linder}, E. and {Lodha}, K. and {Magneville}, C. and {Manera}, M. and {Margala}, D. and {Martini}, P. and {Maus}, M. and {McDonald}, P. and {Medina-Varela}, L. and {Meisner}, A. and {Mena-Fern{\'a}ndez}, J. and {Miquel}, R. and {Moon}, J. and {Moore}, S. and {Moustakas}, J. and {Mueller}, E. and {Mu{\~n}oz-Guti{\'e}rrez}, A. and {Myers}, A.~D. and {Nadathur}, S. and {Napolitano}, L. and {Neveux}, R. and {Newman}, J.~A. and {Nguyen}, N.~M. and {Nie}, J. and {Niz}, G. and {Noriega}, H.~E. and {Padmanabhan}, N. and {Paillas}, E. and {Palanque-Delabrouille}, N. and {Pan}, J. and {Penmetsa}, S. and {Percival}, W.~J. and {Pieri}, M.~M. and {Pinon}, M. and {Poppett}, C. and {Porredon}, A. and {Prada}, F. and {P{\'e}rez-Fern{\'a}ndez}, A. and {P{\'e}rez-R{\`a}fols}, I. and {Rabinowitz}, D. and {Raichoor}, A. and {Ram{\'\i}rez-P{\'e}rez}, C. and {Ramirez-Solano}, S. and {Rashkovetskyi}, M. and {Ravoux}, C. and {Rezaie}, M. and {Rich}, J. and {Rocher}, A. and {Rockosi}, C. and {Roe}, N.~A. and {Rosado-Marin}, A. and {Ross}, A.~J. and {Rossi}, G. and {Ruggeri}, R. and {Ruhlmann-Kleider}, V. and {Samushia}, L. and {Sanchez}, E. and {Saulder}, C. and {Schlafly}, E.~F. and {Schlegel}, D. and {Schubnell}, M. and {Seo}, H. and {Shafieloo}, A. and {Sharples}, R. and {Silber}, J. and {Slosar}, A. and {Smith}, A. and {Sprayberry}, D. and {Tan}, T. and {Tarl{\'e}}, G. and {Taylor}, P. and {Trusov}, S. and {Ure{\~n}a-L{\'o}pez}, L.~A. and {Vaisakh}, R. and {Valcin}, D. and {Valdes}, F. and {Vargas-Maga{\~n}a}, M. and {Verde}, L. and {Walther}, M. and {Wang}, B. and {Wang}, M.~S. and {Weaver}, B.~A. and {Weaverdyck}, N. and {Wechsler}, R.~H. and {Weinberg}, D.~H. and {White}, M. and {Yu}, J. and {Yu}, Y. and {Yuan}, S. and {Y{\`e}che}, C. and {Zaborowski}, E.~A. and {Zarrouk}, P. and {Zhang}, H. and {Zhao}, C. and {Zhao}, R. and {Zhou}, R. and {Zhuang}, T.},
        title = "{DESI 2024 VI: cosmological constraints from the measurements of baryon acoustic oscillations}",
      journal = {\jcap},
         year = 2025,
        month = feb,
       volume = {2025},
       number = {2},
          eid = {021},
        pages = {021},
          doi = {10.1088/1475-7516/2025/02/021},
archivePrefix = {arXiv},
       eprint = {2404.03002},
 primaryClass = {astro-ph.CO},
       adsurl = {https://ui.adsabs.harvard.edu/abs/2025JCAP...02..021A}
}

@ARTICLE{Moreno-Raya16,
       author = {{Moreno-Raya}, Manuel E. and {Moll{\'a}}, Mercedes and {L{\'o}pez-S{\'a}nchez}, {\'A}ngel R. and {Galbany}, Llu{\'\i}s and {V{\'\i}lchez}, Jos{\'e} Manuel and {Carnero Rosell}, Aurelio and {Dom{\'\i}nguez}, Inmaculada},
        title = "{On the Dependence of  Type Ia SNe Luminosities on the Metallicity of Their Host Galaxies}",
      journal = {\apjl},
         year = 2016,
        month = feb,
       volume = {818},
       number = {1},
          eid = {L19},
        pages = {L19},
          doi = {10.3847/2041-8205/818/1/L19},
archivePrefix = {arXiv},
       eprint = {1511.05348},
 primaryClass = {astro-ph.GA},
       adsurl = {https://ui.adsabs.harvard.edu/abs/2016ApJ...818L..19M}
}

@ARTICLE{Shuntov25,
       author = {{Shuntov}, Marko and {Akins}, Hollis B. and {Paquereau}, Louise and {Casey}, Caitlin M. and {Ilbert}, Olivier and {Arango-Toro}, Rafael C. and {McCracken}, Henry Joy and {Franco}, Maximilien and {Harish}, Santosh and {Kartaltepe}, Jeyhan S. and {Koekemoer}, Anton M. and {Yang}, Lilan and {Huertas-Company}, Marc and {Berman}, Edward M. and {McCleary}, Jacqueline E. and {Toft}, Sune and {Gavazzi}, Rapha{\"e}l and {Achenbach}, Mark J. and {Bertin}, Emmanuel and {Brinch}, Malte and {Champagne}, Jackie and {Chartab}, Nima and {Drakos}, Nicole E. and {Egami}, Eiichi and {Endsley}, Ryan and {Faisst}, Andreas L. and {Fan}, Xiaohui and {Flayhart}, Carter and {Hartley}, William G. and {Hatamnia}, Hossein and {Gozaliasl}, Ghassem and {Gentile}, Fabrizio and {Jermann}, Iris and {Jin}, Shuowen and {Kakiichi}, Koki and {Khostovan}, Ali Ahmad and {K{\"u}mmel}, Martin and {Laigle}, Clotilde and {Laishram}, Ronaldo and {Lambrides}, Erini and {Liu}, Daizhong and {Lyu}, Jianwei and {Magdis}, Georgios and {Mobasher}, Bahram and {Moutard}, Thibaud and {Renzini}, Alvio and {Robertson}, Brant E. and {Schefer}, Marc and {Scognamiglio}, Diana and {Scoville}, Nick and {Sattari}, Zahra and {Sanders}, David B. and {Taamoli}, Sina and {Trakhtenbrot}, Benny and {Valentino}, Francesco and {Wang}, Feige and {Weaver}, John R. and {Yang}, Jinyl},
        title = "{COSMOS2025: The COSMOS-Web galaxy catalog of photometry, morphology, redshifts, and physical parameters from JWST, HST, and ground-based imaging}",
      journal = {arXiv e-prints},
         year = 2025,
        month = jun,
          eid = {arXiv:2506.03243},
        pages = {arXiv:2506.03243},
          doi = {10.48550/arXiv.2506.03243},
archivePrefix = {arXiv},
       eprint = {2506.03243},
 primaryClass = {astro-ph.GA},
       adsurl = {https://ui.adsabs.harvard.edu/abs/2025arXiv250603243S}
}

@ARTICLE{Dettman21,
       author = {{Dettman}, Kyle G. and {Jha}, Saurabh W. and {Dai}, Mi and {Foley}, Ryan J. and {Rest}, Armin and {Scolnic}, Daniel M. and {Siebert}, Matthew R. and {Chambers}, K.~C. and {Coulter}, D.~A. and {Huber}, M.~E. and {Johnson}, E. and {Jones}, D.~O. and {Kilpatrick}, C.~D. and {Kirshner}, R.~P. and {Pan}, Y. -C. and {Riess}, A.~G. and {Shultz}, A.~S.~B.},
        title = "{The Foundation Supernova Survey: Photospheric Velocity Correlations in Type Ia Supernovae}",
      journal = {\apj},
         year = 2021,
        month = dec,
       volume = {923},
       number = {2},
          eid = {267},
        pages = {267},
          doi = {10.3847/1538-4357/ac2ee5},
archivePrefix = {arXiv},
       eprint = {2102.06524},
 primaryClass = {astro-ph.HE},
       adsurl = {https://ui.adsabs.harvard.edu/abs/2021ApJ...923..267D}
}

@ARTICLE{Pan24,
       author = {{Pan}, Y. -C. and {Jheng}, Y. -S. and {Jones}, D.~O. and {Lee}, I. -Y. and {Foley}, R.~J. and {Chornock}, R. and {Scolnic}, D.~M. and {Berger}, E. and {Challis}, P.~M. and {Drout}, M. and {Huber}, M.~E. and {Kirshner}, R.~P. and {Kotak}, R. and {Lunnan}, R. and {Narayan}, G. and {Rest}, A. and {Rodney}, S. and {Smartt}, S.},
        title = "{Measuring the ejecta velocities of type Ia supernovae from the pan-STARRS1 medium deep survey}",
      journal = {\mnras},
         year = 2024,
        month = aug,
       volume = {532},
       number = {2},
        pages = {1887-1900},
          doi = {10.1093/mnras/stae1618},
archivePrefix = {arXiv},
       eprint = {2211.06895},
 primaryClass = {astro-ph.HE},
       adsurl = {https://ui.adsabs.harvard.edu/abs/2024MNRAS.532.1887P}
}

@ARTICLE{Gall24,
       author = {{Gall}, Christa and {Izzo}, Luca and {Wojtak}, Radoslaw and {Hjorth}, Jens},
        title = "{The Hubble Constant from Blue Type Ia Supernovae}",
      journal = {arXiv e-prints},
         year = 2024,
        month = nov,
          eid = {arXiv:2411.05642},
        pages = {arXiv:2411.05642},
          doi = {10.48550/arXiv.2411.05642},
archivePrefix = {arXiv},
       eprint = {2411.05642},
 primaryClass = {astro-ph.CO},
       adsurl = {https://ui.adsabs.harvard.edu/abs/2024arXiv241105642G}
}
\bibliographystyle{aasjournal}

\end{document}